\documentclass[journal,twoside,web]{ieeecolor}
\usepackage{generic}

\usepackage{amssymb,amsmath,latexsym,amsfonts,amsthm,mathtools}

\usepackage{cite,graphicx}
\usepackage{float,subcaption}
\usepackage{hyperref}
\hypersetup{hidelinks}
\usepackage{textcomp}
\newtheorem{theorem}{Theorem}
\newtheorem{lemma}{Lemma}

\newtheorem{problem}{Problem}

\theoremstyle{remark}
\newtheorem{remark}{Remark}
\newtheorem{assumption}{Assumption}

\theoremstyle{definition}

\usepackage{anyfontsize}

\usepackage{newtxmath}
\usepackage{booktabs}
\allowdisplaybreaks

\def\BibTeX{{\rm B\kern-.05em{\sc i\kern-.025em b}\kern-.08em
    T\kern-.1667em\lower.7ex\hbox{E}\kern-.125emX}}
\begin{document}
\title{Provably Safe Control for Constrained Nonlinear Systems with Bounded Input}
\author{Saurabh Kumar, Shashi Ranjan Kumar, \IEEEmembership{Senior Member, IEEE}, and Abhinav Sinha, \IEEEmembership{Senior Member, IEEE}
\thanks{S. Kumar and S.R. Kumar are with the Intelligent Systems and Control (ISaC) Lab, Department of Aerospace Engineering, Indian Institute of Technology Bombay, Maharashtra 400076 India (e-mails: \{saurabh.k,srk\}@aero.iitb.ac.in). }
\thanks{A. Sinha is with the Guidance, Autonomy, Learning, and Control for Intelligent Systems (GALACxIS) Lab, Department of Aerospace Engineering and Engineering Mechanics, University of Cincinnati, OH 45221, USA. (e-mail: abhinav.sinha@uc.edu).}}

\maketitle
\begin{abstract}
In real-world control applications, actuator constraints and output constraints (specifically in tracking problems) are inherent and critical to ensuring safe and reliable operation. However, generally, control strategies often neglect these physical limitations, leading to potential instability, degraded performance, or even system failure when deployed on real-world systems. This paper addresses the control design problem for a class of nonlinear systems under both actuator saturation and output constraints. First, a smooth asymmetric saturation model (a more generic representative of practical scenarios) is proposed to model actuator saturation, which ensures that the control inputs always remain confined within a predefined set to ensure safety. Based on the proposed model, we develop a nonlinear control framework that guarantees output tracking while ensuring that system output remains confined to the predefined set. Later, we integrate this design with the constrained output tracking control problem, wherein we show that the system output tracks its desired trajectory by simultaneously satisfying input and output constraints. The global stabilization of the tracking error is achieved in the presence of input constraints, while semi-global stabilization is achieved in the presence of both input and output constraints. Additionally, we rigorously establish the boundedness of all closed-loop signals under the proposed design. Simulation results demonstrate the effectiveness of the proposed methods in handling asymmetric constraints while achieving desirable tracking performance. %
\end{abstract}

\begin{IEEEkeywords}
Asymmetric saturation model, input constraint, output constraint, bounded nonlinear control, safety.
\end{IEEEkeywords}
\section{Introduction}
\label{sec:introduction}
\IEEEPARstart{D}{espite} significant advancements in control systems, their practical application is often hindered by physical constraints inherent in real-world systems. These constraints are pervasive in control systems, and one of the most common constraints pertains to actuators. The actuator constraints stem from the fact that most practical systems are driven by actuators, which always have finite capabilities. Although input constraints are invariably present in practice, they are frequently neglected in the controller design. However, ignoring the physical capabilities of the actuators may be detrimental to the stability and performance of the control systems and may even drive the system into an unsafe mode, potentially causing damage to the actuators. Therefore, for a system's safe and reliable operation, the actuator saturation should be accounted for in the controller's design. 

We can broadly categorize the existing research on control design considering input saturation into two categories: global stabilization and semi-global stabilization strategies. In semi-global stabilization, the initial states of the system are constrained to lie within a specified region of the state space, whereas global stabilization imposes no such restriction. A pivotal result on the stabilization of linear systems subjected to bounded control was presented in \cite{203432}, where it was shown that a linear system can be globally asymptotically stabilized under bounded inputs if and only if the open-loop system is asymptotically null controllable\footnotemark. The works in \cite{doi:10.1080/00207176908905846,261255} highlighted the limitations of linear feedback controllers in achieving global stabilization for systems with more than two integrators. Specifically, it was shown in \cite{261255} that a system of $n$ integrators with $n \geq 3$ cannot be globally stabilized using linear feedback with bounded control. \footnotetext{A linear system ($\dot{\mathbf{x}} =  \mathbf{Ax} + \mathbf{Bu}$) is said to be asymptotically null controllable if and only if (i) no eigenvalues of $\mathbf{A}$ are in the right half of $s$-plane (ii) The pair ($\mathbf{A}, \mathbf{B}$) is stabilizable in the ordinary sense, that is, all the uncontrollable modes of the system have negative real parts.}

Since linear controllers are unable to guarantee the global stabilization of such systems with bounded control input, various methods were presented in \cite{doi:10.1016/0167-6911(93)90033-3,doi:10.1002/rnc.4590050503,486638,4610041} to achieve semi-global stabilization. In \cite{doi:10.1016/0167-6911(93)90033-3}, the authors used a low-gain feedback design to achieve semi-global stabilization with a bounded control input. The key idea of this design was to choose as small feedback gains as possible. However, the system's dynamic performance may be compromised as the actuators' control ability is not fully utilized. Later, to utilize the actuators' full capability, high-low gain concepts were introduced in \cite{doi:10.1002/rnc.4590050503} and later completed in \cite{486638}. This method superimposed high- and low-gain controllers, each independently tunable, allowing improved actuator utilization and an enlarged domain of attraction. This method was a superposition of high and low gain controllers, both of which can be independently designed with adjustable parameters. By adjusting those design parameters, the actuator utilization was improved, which consequently enhanced the domain of attraction of the closed-loop system. However, this method remains limited in enhancing the dynamic performance of the closed-loop system. An alternative approach to the low gain method, based on the solution of a parametric Lyapunov equation, is presented in \cite{4610041}. In order to improve the dynamic performance of the closed-loop system, a time-varying state feedback controller and an observer-based time-varying output feedback controller were designed in \cite{doi:10.1016/j.jfranklin.2020.08.025} using the parametric Riccati and Lyapunov equations. 

For global stabilization of a chain of integrators, a nested saturation-based feedback control was proposed in \cite{doi:10.1016/0167-6911(92)90001-9}, which was later generalized in \cite{362853}. Various adaptive control techniques were also discussed in \cite{doi:10.1016/j.automatica.2005.02.009,728887,doi:10.1016/S0005-1098(00)00154-0,doi:10.1016/0005-1098(94)90202-X,333787,doi:10.1016/0005-1098(95)00059-6} for handling input constraints. The strategies in \cite{doi:10.1016/j.automatica.2005.02.009,doi:10.1016/0005-1098(94)90202-X}, initially designed controllers without considering the effect of input saturation and later added a compensation term to mitigate the adverse effect on the closed-loop performance caused by the input saturation. The strategies in \cite{doi:10.1016/S0005-1098(00)00154-0,333787,doi:10.1016/0005-1098(95)00059-6} considered the input saturation as the discrepancy between controller output and internal controller states, which was corrected through modifying the controller inputs. Note that these control techniques were primarily developed for linear systems. Since most of the real-world systems exhibit nonlinear dynamics, the development of nonlinear controllers that can accommodate actuator saturation is imperative. For instance, \cite{5723705} used a combination of hyperbolic tangent and \textit{Nussbaum} functions to deal with the problem of input saturation and then designed an adaptive backstepping-based controller for output tracking of a nonlinear system. The design was later extended in \cite{6975243} for a class of uncertain nonlinear systems, employing radial basis function neural networks to approximate unknown nonlinearities and a nonlinear disturbance observer to handle the external disturbances. 


It is important to note that the majority of the above-discussed methods assume a symmetric input saturation model. While the symmetric saturation model is a degenerate case of the asymmetric saturation model, its direct application to asymmetric saturation problems yields conservative solutions. As highlighted in the recent survey paper \cite{doi:10.1080/00207721.2021.1989726},  consideration of the asymmetric saturation model is crucial and has more practical significance. However, to the best of the authors' knowledge, the explicit treatment of asymmetric input saturation for nonlinear systems remains largely unexplored in the literature. There are only a few works, for example, \cite{6875955,7428909,6994268} reported in the literature that deal with handling asymmetric actuator saturation. The work in \cite{6875955} used a Gaussian error function to model asymmetric actuator constraints. In \cite{7428909}, neural networks and adaptive control were used for tracking problems in the presence of asymmetric input constraints, while in \cite{6994268}, it was used for a non-affine nonlinear system. The key idea of these designs was to approximate unknown functions using a neural network and utilize an auxiliary system to analyze the effect of input saturation. 

In the light of the above-mentioned literature, the main contributions of this work are fourfold.
\begin{itemize}
    \item We propose a smooth asymmetric saturation model that enforces strict input constraints, ensuring all control signals remain within predefined safe bounds while preserving closed-loop stability, thereby introducing safety in the control design.
    \item We augment the proposed model with a nonlinear plant to ensure exact output tracking under bounded control inputs. Through rigorous analysis, we show that the tracking error dynamics admit global stabilization despite input constraints.
    \item Moreover, we investigate the constrained output tracking problem subject to bounded control input constraints, establishing semi-global stabilizability of the associated error dynamics.
    \item We provide rigorous proof for the uniform boundedness of all closed-loop signals under the proposed control scheme.
\end{itemize}
The rest of the paper is organized as follows. After a brief description of the existing works, we formulate the problems in Section~\ref{sec:probelm_formulation}. We design the safe nonlinear controllers in Section~\ref{sec:main_results} followed by the validations of the proposed controller in Section~\ref{sec:simulations}. We provide concluding remarks in Section~\ref{sec:conclusions} with some interesting future directions.

\section{Problem Formulation}\label{sec:probelm_formulation}
Consider a nonlinear system in the strict feedback form as
\begin{subequations}\label{eq:system_dynamics}
    \begin{align}
        \dot{x}_{i} =&~f_{i}\left(\bar{x}_{i}\right) + g_{i}\left(\bar{x}_{i}\right) x_{i+1}, \quad i = 1, 2, \ldots, n-1, \label{eq:xi_dot}\\
         \dot{x}_{n} =&~f_{n}\left(\bar{x}_{n}\right) + g_{n}\left(\bar{x}_{n}\right) u , \label{eq:xn_dot}\\
         y =&~ x_{1}, \label{eq:y}
    \end{align}
\end{subequations}
where, $\forall~i$, $f_{i}\left(\bar{x}_{i}\right)$, $g_{i}\left(\bar{x}_{i}\right)$ $:\mathbb{R}^{i} \to \mathbb{R}$ are smooth nonlinear functions, the $n$-tuple ($x_1$, $x_2$, $\cdots, x_n$) $\in \mathbb{R}^n$ and $y\in\mathbb{R}$ denote states and output of the system, $\bar{x}_{i} = [x_1, x_2, x_3, \cdots, x_i]^\top \in \mathbb{R}^{i}$ is the state vector, and $u \in \mathbb{R}$ is the control input to the system. 

We consider that the plant is subject to an asymmetric saturation type nonlinearity in which the plant input can be described by $u(v)$, where $v$ is a representative command signal that needs to be designed. In other words, the plant input $u$ is also the input to its actuator whose safe operation is restricted to an open set $(u_{\min},u_{\max})$, where $u_{\min}< 0$ and $u_{\max}>0$ are lower and upper bounds of $u(t)$ in the proposed saturation model. The asymmetric saturation for some signal $v$ can be expressed as
\begin{equation} \label{eq:sat_def}
u(v) = \mathrm{sat}(v)= \begin{cases}
                            u_{\max}; & \text{if}\; v > u_{\max},\\
                            v ; & \text{if}\; u_{\min} \leq v \leq u_{\max},\\
                           u_{\min}; & \text{if}\; v < u_{\min}, 
                            \end{cases} 
\end{equation}
One can observe from \eqref{eq:sat_def} that the relationship between the plant input $u(t)$ and the controller output $v(t)$ exhibits a sharp corner point whenever the controller output hits the saturation boundaries, that is $v(t) = u_{\min}$ and $v(t)=u_{\max}$. Consequently, the \emph{backstepping} control techniques cannot be directly applied. 
\begin{problem}[Output tracking problem with bounded control]\label{problem1}
 For the nonlinear system given in \eqref{eq:system_dynamics},  design a smooth input saturation model such that
 \begin{itemize}
     \item The system output tracks the desired trajectory$\colon$
   $$\lim_{t \to \infty} \|y(t) - y_d(t)\| = 0,$$
   \item  The control input $u(t)$ satisfies the asymmetric actuator constraint for all $t \geq 0\colon$
   $$u(t) \in \mathbb{U} \coloneq \{ u \in \mathbb{R}~ \rvert u_{\min} < u < u_{\max} \}$$
 \end{itemize}
\end{problem}
In other words, we aim to develop a smooth input saturation model such that the system output $y$ tracks its desired trajectory $y_{d}(t)$ while respecting the asymmetric actuator constraints.  Additionally, in certain applications, it may be desirable to guarantee that the system's output also remains confined to a predefined set for a safe and reliable operation. So, we state our next problems as follows. 
\begin{problem}[Constrained output tracking with bounded control]\label{problem2}
Consider the nonlinear system given in \eqref{eq:system_dynamics}. Design a nonlinear control strategy such that
\begin{itemize}
    \item Output tracking is achieved:
    $$\lim_{t \to \infty} \|y(t) - y_d(t)\| = 0,$$
    \item Output remains within a prescribed time-varying safe set:
    $$y(t) \in \mathbb{Y} := \{ y \in \mathbb{R}~\rvert \underline{y}(t) < y(t) < \overline{y}(t) \}, \quad \forall t \geq 0,$$
     where $\underline{y}(t), \overline{y}(t): \mathbb{R}_+ \to \mathbb{R}$ are continuous functions satisfying  $\overline{y}(t) \geq \underline{y}(t)\; \forall t \geq 0,$
    \item Actuator constraints are respected:
    $$u(t) \in \mathbb{U} := \{ u \in \mathbb{R} ~\rvert u_{\min} < u < u_{\max} \}, \quad \forall t \geq 0.$$
\end{itemize}
\end{problem}
\begin{remark}
    We proceed with the control design under the standard assumption that \eqref{eq:system_dynamics} is input-to-state (ISS) stable and possesses sufficient control authority for stabilization under input constraints, which is physically justified. For example, autonomous vehicles admit stable trajectory tracking through proper control synthesis despite operating under strict bounds on thrust and angular rates. For any ISS system with well-defined input constraints, there exists a nonempty set of feasible control signals capable of achieving the desired behavior without violating physical constraints. Furthermore, systems lacking inherent stability or requiring unrealizable control efforts fundamentally cannot achieve arbitrary performance specifications under bounded inputs.
\end{remark}
Before proceeding to formally design the control strategies, we present some lemmas that will aid in the design.
\begin{lemma}(\cite{doi:10.1016/j.automatica.2008.11.017}) \label{lem:BLF}
    Let $\mathbb{T}:=\{ \chi \in \mathbb{R}:\lvert \chi \rvert < 1 \} \subset \mathbb{R}$ and $\mathbb{Q}:=\mathbb{R}^{\ell} \times \mathbb{T} \subset \mathbb{R}^{\ell+1}$ be open sets. Consider the system $\dot{\lambda}=f(t,\lambda)$, where $\lambda=[\omega, \chi]^\top \in \mathbb{Q}$, and $f:\mathbb{R}_{+} \times \mathbb{Q} \to \mathbb{R}^{\ell+1}$ is locally Lipschitz  in $\omega$ and piecewise continuous and uniformity in $t$ on $\mathbb{R}_{+} \times \mathbb{Q}$. Suppose that there exist functions $\Gamma_1:\mathbb{R}^{\ell} \times \mathbb{R}_+ \to \mathbb{R}_{+} $ and $\Gamma_2:\mathbb{T} \to \mathbb{R}_{+} $ positive definite and continuously differentiable in their respective domains such that the following holds $\Gamma_2(\chi) \to \infty$ as $\chi \to \infty $ and $\mathscr{K}_1\left( \|\omega\| \right) \leq \Gamma_1 \left( \omega,t \right) \leq \mathscr{K}_2\left( \|\omega\| \right)$, where $\mathscr{K}_1,\mathscr{K}_2$ belong to the class of $\mathscr{K}_\infty$ functions. Let $\Gamma=\Gamma_2(\chi)+\Gamma_1(\omega,t)$, and $\chi(0)=0 \in \mathbb{T}$. If the inequality $\dot{\Gamma}=\frac{\partial \Gamma}{\partial \lambda} f \leq 0$ in the set $\chi \in \mathbb{T}$, then $\chi(t) \in \mathbb{T}$ for all time $t \geq 0$.
\end{lemma}
\begin{lemma} (\cite{doi:10.1016/j.automatica.2008.11.017})   \label{eq:blf_log_inequality}
    The inequality $\ln\left( \frac{1}{1-\chi^{2k}} \right) < \frac{\chi^{2k}}{1-\chi^{2k}}$ hold for $\lvert \chi \rvert < 1$ and $k\in \mathbb{N}$, where $\mathbb{N}$ denotes the set of natural numbers.
\end{lemma}
\section{Main Results}\label{sec:main_results}
In this section, we first propose a smooth input saturation model that accounts for the asymmetric actuator constraints. Thereafter, we augment the proposed model with the nonlinear plant considered in \eqref{eq:system_dynamics} to achieve the objective of global output tracking with bounded input. Subsequently, we consider the problem of constrained output tracking and augment the proposed model to achieve the objective of semi-global output tracking with bounded input. We begin the controller design by presenting the proposed saturation model in the following subsection. For brevity, we drop the arguments denoting the time dependency of variables if no confusion arises.

\subsection{Asymmetric Input Saturation Model}
\begin{figure}[!ht]
    \centering
    \includegraphics[width=\linewidth]{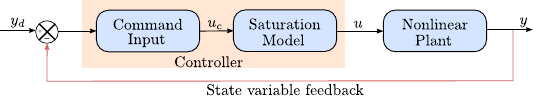}
    \caption{A depiction of bounded input scenario.}
    \label{fig:saturation_model_block}
\end{figure}

Instead of using the widespread saturation model given in \eqref{eq:sat_def}, which has a sharp corner point in the control input at the boundary, we propose a smooth asymmetric nonlinear actuator saturation model as
\begin{align} \label{eq:u_saturation_model}
\dot{u} =& p_1 \left[\rho  \left\{ 1 - \left(\dfrac{u}{u_{\max}}\right)^{\gamma} \right\}  + \left(1 - \rho \right)  \left\{ 1 - \left(\dfrac{u}{u_{\min}}\right)^{\gamma} \right\}  \right] u_{c} \nonumber\\
&-p_1p_2 u,
\end{align}
with $u(0)=0$, where $u_{c}$ is the model input (commanded input; refer Fig.~\ref{fig:saturation_model_block}), $u$ is the model output (control input to the plant), $p_1,p_2 \in\mathbb{R}_+$ are constants, $\gamma \geq 2n$ such that $n \in \mathbb{N}$, $u_{\min}$ and $u_{\max}$ are minimum and maximum available control inputs to the system such that $u_{\min} < 0$ and $u_{\max} > 0$. The term $\rho$ is defined as $\rho =1$ if $u>0$ and $\rho=0$ if $u \leq 0$. Note that under the proposed model, both $u_{\max}$ and $u_{\min}$ cannot be zero at the same time to ensure that the model is well-defined.
\begin{remark}
By changing the values of $u_{\max}$ and $u_{\min}$, the bounds on the actuator magnitude can be easily adjusted according to the design requirements. For a choice $u_{\min} = u_{\max}$, the symmetric bounds can be obtained while $u_{\min} \neq  u_{\max}$ in \eqref{eq:u_saturation_model} represents an asymmetric saturation model. Consequently, the proposed model provides a degree of generalizability to the design.
\end{remark}

\begin{theorem}\label{thm:sat_model}
Consider the saturation model in \eqref{eq:u_saturation_model}. If the commanded input, $u_{c}$, remains finite for all $t \geq 0$, then the plant input $u$ remains confined to the set $\mathbb{U}$, which is defined as $\mathbb{U} \coloneq \{u\, \rvert\, u_{\min} < u(t) < u_{\max} \}$ $\forall$ $t \geq 0$.
\end{theorem}
\begin{proof}
Let the commanded input $u_{c}$ be finite, i.e., there exists a constant $\xi > 0$ such that $|u_{c}| \leq \xi < \infty$ for all $t \geq 0$. We show that the system input $u$ (which is the model output) remains strictly within the bounds $u_{\min}$ and $u_{\max}$ by splitting the analysis in the proof into two regions, viz. $u>0$ and $u<0$. 

\textit{Case I}: Consider the case where $u > 0$. If $u = u_{\max}$, then from \eqref{eq:u_saturation_model}, we have
\begin{equation*}
    \dot{u} = - p_1 p_2 u_{\max}.
\end{equation*}
Since $p_1,p_2\in\mathbb{R}_+$, it follows that $\dot{u} < 0$. This implies that if $u$ reaches $u_{\max}$, it will start decreasing. Thus, $u$ cannot stay at or exceed $u_{\max}$. Now, we aim to show that that $u$ will never reach its maximum value $u_{\max}$. To further refine our argument, consider the general case when $u\in(u, u_{\max})$. From \eqref{eq:u_saturation_model}, we can write
\begin{equation} \label{eq:upper_bound_derivative}
    \dot{u} = p_1 \left[ 1 - \left( \dfrac{u}{u_{\max}} \right)^{\gamma} \right] u_c - p_1 p_2 u.
\end{equation}
Since $1 - (u/u_{\max})^{\gamma} \geq 0$, $\dot{u}$ can increase only if $u_c > 0$. Without loss of generality, let us assume that $u_{c} > 0$, then using the fact that $|u_c| \leq \xi$, one may write
\begin{align}
    \dot{u} \leq &~ p_1 \left[ 1 - \left(\dfrac{u}{u_{\max}}\right)^{\gamma} \right] \xi - p_1 p_2 u \nonumber\\
    \leq&~ p_1 \left[ 1 - \left(\dfrac{u}{u_{\max}}\right)^{\gamma} -\dfrac{p_2 u_{\max} }{\xi} \left( \dfrac{u}{u_{\max}}\right) \right] \xi, \label{eq:upper_bound_final}
\end{align}
which may be written in a more convenient form 
\begin{equation}\label{eq:vu_saturation_model_3}
    \dot{u} \leq  p_1 \xi \left[1 -\left(1 +\dfrac{p_2 u_{\max} }{\xi} \right) \left(\dfrac{u}{u_{\max}} \right)^{\gamma} \right]
\end{equation}
since $-\dfrac{u}{u_{\max}} \leq -\left( \dfrac{u}{u_{\max}} \right)^{\gamma}$.

One may observe from \eqref{eq:vu_saturation_model_3} that $\dot{u} \leq 0$ if
\begin{align}
    \left[ 1 -\left(1 +\frac{p_2 u_{\max} }{\xi} \right) \left(\frac{u}{u_{\max}} \right)^{\gamma}\right] \leq 0,
\end{align}
or, equivalently,
\begin{equation*}
\left(1 +\dfrac{p_2 u_{\max} }{\xi} \right) \left(\dfrac{u}{u_{\max}} \right)^{\gamma} \geq 1,
\end{equation*}
which, on solving for $u$, yields 
\begin{equation}
u \geq u_{\max} \left[ \dfrac{\xi}{\xi + p_2u_{\max}}  \right] ^{1/\gamma} \coloneqq \tilde{u}_{\max}\;(\text{say}).
\end{equation}
If $u \geq \tilde{u}_{\max}$, then the term in brackets in \eqref{eq:upper_bound_final} is non-positive, implying $\dot{u} \leq 0$. Since $\tilde{u}_{\max} < u_{\max}$, this guarantees that $u$ will always remain below $u_{\max}$.

\textit{Case II}: Now, consider $u < 0$. If $u = u_{\min}$, then from \eqref{eq:u_saturation_model}, we have
\begin{equation*}
    \dot{u} = - p_1 p_2 u_{\min}.
\end{equation*}
Since $u_{\min} < 0$ and $p_1, p_2 \in \mathbb{R}_+$, we get $\dot{u} > 0$, implying that $u$ will increase if it reaches $u_{\min}$, preventing it from decreasing further. For the general case where $u \in (u_{\min} , 0]$, one may write \eqref{eq:u_saturation_model} as
\begin{equation} \label{eq:lower_bound_derivative}
    \dot{u} = p_1 \left[ 1 - \left( \frac{u}{u_{\min}} \right)^{\gamma} \right] u_{c} + p_1 p_2 u.
\end{equation}
One may notice that $\dot{u}$ can decrease only if $u_c < 0$ as the term $ 1 - \left(\dfrac{u}{u_{\min}}\right)^{\gamma} \geq 0\,\forall\,u\leq 0$. 

Without loss of generality, let us consider $u_c < 0$, which allows to write \eqref{eq:u_saturation_model}, using $-\xi \leq u_{c}$, as
\begin{align}
    \dot{u} \geq&~  p_1  \left[ 1 - \left(\dfrac{u}{u_{\min}}\right)^{\gamma} \right] \left(-\xi\right)  + p_1p_2 u, \nonumber \\
    \geq&~ p_1 \xi \left[ -1 + \left(\dfrac{u}{u_{\min}}\right)^{\gamma} + \dfrac{p_2 u_{\min} }{\xi} \left( \dfrac{u}{u_{\min}}\right) \right] ,
\end{align}
which may be simplified to
\begin{equation} \label{eq:lower_bound_final}
    \dot{u} \geq p_1 \xi \left[ -1 + \left( 1 + \dfrac{p_2 u_{\min}}{\xi_M} \right) \left( \dfrac{u}{u_{\min}} \right)^{\gamma} \right].
\end{equation}
upon noting that $\dfrac{u}{u_{\min}} \geq \left( \dfrac{u}{u_{\min}} \right)^{\gamma}$.

Observe from \eqref{eq:lower_bound_final} that $\dot{u} \geq 0$ if the inequality 
\begin{equation*}
-1  + \left(1 + \dfrac{p_2 u_{\min} }{\xi} \right) \left(\dfrac{u}{u_{\min}} \right)^{\gamma} \geq 0
\end{equation*}
holds true, which points out to fact that $\dot{u} \geq 0$ if
\begin{align}
\left(\dfrac{u}{u_{\min}} \right)^{\gamma} \left(\dfrac{\xi + p_2u_{\min}}{\xi}\right)  \geq&~ 1, 
\end{align}
which implies
\begin{align}
    u \geq u_{\min} \left[ \dfrac{\xi}{\xi + p_2u_{\min}}  \right] ^{1/\gamma}&~ := \tilde{u}_{\min}\;(\text{say}). \label{eq:vu_saturation_model_4_lower}
\end{align}
If $u \leq \tilde{u}_{\min}$ then the term in brackets in \eqref{eq:lower_bound_final} is non-negative, implying $\dot{u} \geq 0$. Since $\tilde{u}_{\min} > u_{\min}$, this confirms that $u$ will always remain above $u_{\min}$.
Therefore, we have established that there exist constants $\tilde{u}_{\max}$ and $\tilde{u}_{\min}$ such that $\tilde{u}_{\min} < u <\tilde{u}_{\max}$, where $u_{\min} < \tilde{u}_{\min} \leq u \leq \tilde{u}_{\max} < u_{\max}$
Thus, the plant input $u$ strictly remains confined to the set $\mathbb{U}\;\forall\,t \geq 0$.
\end{proof}

\begin{remark}
Theorem~\ref{thm:sat_model} essentially infers that if a commanded input can be designed to remain bounded (even if the exact bounds are not explicitly known), the model output (the actual input to the system) will be guaranteed to be confined within an asymmetric bound. 
\end{remark}
\subsection{Output Tracking with Bounded Control}
In this subsection, we augment the proposed model with the nonlinear plant given in \eqref{eq:system_dynamics} to solve Problem~\ref{problem1}. The augmented system can be obtained as
\begin{subequations}\label{eq:system_dynamics_aug_global}
    \begin{align}
        \dot{x}_{i} =&~f_{i}\left(\bar{x}_{i}\right) + g_{i}\left(\bar{x}_{i}\right) x_{i+1}, \quad i = 1, 2, \cdots n-1 \label{eq:xi_dotaug_global}\\
         \dot{x}_{n} =&~f_{n}\left(\bar{x}_{n}\right) + g_{n}\left(\bar{x}_{n}\right) u , \label{eq:xn_dotaug_global}\\
         \dot{u} =&~g(u) u^c - f(u) \label{eq:u_dotaug_global}\\
         y =&~ x_{1}, \label{eq:y_aug_global}
    \end{align}
\end{subequations}
Note that the proposed model \eqref{eq:u_saturation_model} can be considered as a part of the control strategy in which the model output is the input to the nonlinear plant \eqref{eq:system_dynamics}. Now, if we can design any commanded input that is finite, then the model output (or input to the plant) will always remain confined to set $\mathbb{U}$, and consequently, the overloading of the actuators can be avoided. For the sake of controller design, the following assumptions are needed.
\begin{assumption}
The desired trajectory $y_d$ and its derivative up to order $n+1$ are continuously bounded.
\end{assumption}
\begin{assumption}\label{assum:1}
The signs of the functions $g_{i}\left(\bar{x}_{i}\right)$ for $i=1,2,\cdots,n$ are known and there exists a constant $\overline{g}_{i}\geq \underline{g}_{i} >0$ such that $\underline{g}_{i}\leq \lvert g_{i}\left(\bar{x}_{i}\right) \rvert \leq  \overline{g}_{i}$ for $y=x_{1}$.
\end{assumption}
Assumption~\ref{assum:1} essentially implies that smooth functions $g_{i}\left(\bar{x}_{i}\right)$ are strictly positive.

As discussed earlier, the proposed input saturation model does not exhibit sharp corners in the control inputs. Thus, we use the \emph{backstepping} method to design the control input. As in the usual output tracking problem with \emph{backstepping}, the following change of coordinates is made: $\varphi_{1} \coloneqq y - y_{d}$, $\varphi_{i} = x_{i} - \eta_{i-1}$ for $i=2,3,\cdots,n$, where $\eta_{i-1}$ is the stabilizing function for the $i\textsuperscript{th}$ step that needs to be designed, and $\varrho \coloneqq u - \eta_{n}$, where $\varrho$ is an additional state introduced to handle the input saturation and $\eta_{n}$ is the stabilizing function for $n\textsuperscript{th}$ step. To illustrate the \emph{backstepping} procedures, we only elaborate on the details of $1\textsuperscript{st}$, $i\textsuperscript{th}$, and the last two steps, especially to design the commanded control input $u_{c}$.

\textit{Step 1}: Define output tracking error as $\varphi_{1} \coloneqq y - y_{d}$. One can obtain the time derivative of $\varphi_1$ using \eqref{eq:xi_dot} as
\begin{equation} \label{eq:varphi1_dot}
\dot{\varphi}_{1} = \dot{y}- \dot{y}_{d}= \dot{x}_{1}- \dot{y}_{d}= f_{1}(x_1) + g_1(x_{1}) x _{2}- \dot{y}_{d}.
\end{equation}
By viewing $x_{2}$ as a virtual control input, let us consider a Lyapunov function candidate $\mathcal{V}_{1}$ as $\mathcal{V}_{1} = 0.5\varphi_{1}^2$, whose time derivative along the state trajectories results in $\dot{\mathcal{V}}_{1}=\varphi_{1}\dot{\varphi}_{1}$, which by substituting the value of $\dot{\varphi}_{1}$ from \eqref{eq:varphi1_dot}, becomes
\begin{equation} \label{eq:v1_dot_1}
\dot{\mathcal{V}}_{1} = \varphi_{1}\left[ f_{1}(x_1) + g_1(x_{1}) x _{2}- \dot{y}_{d} \right].
\end{equation}
Now, define $\varphi_{2} \coloneqq x_{2} - \eta_{1}$, where $\eta_{1}$ is the stabilizing function. Using the relation $ x_{2} = \varphi_{2} +\eta_{1}$, the expression in \eqref{eq:v1_dot_1} becomes
\begin{equation} \label{eq:v1_dot_2}
\dot{\mathcal{V}}_{1} = \varphi_{1}\left[ f_{1}(x_1) + g_1(x_{1}) \left(\varphi_{2} +\eta_{1} \right)- \dot{y}_{d} \right].
\end{equation}
If we design the stabilizing function $\eta_{1}$ as
\begin{equation} \label{eq:eta_1}
\eta_{1} = \dfrac{1}{g_1(x_{1})}\left[ \dot{y}_{d} - f_{1}(x_1) - k_{1} \varphi_{1}\right],
\end{equation}
then the derivative of the Lyapunov function candidate, $\mathcal{V}_{1}$, becomes 
\begin{equation} \label{eq:v1_dot_final}
\dot{\mathcal{V}}_{1} = -k_{1} \varphi_{1}^{2} +  g_1(x_{1}) \varphi_{2} .
\end{equation}

\textit{Step i ($2 \leq i \leq n-1$)}: In the $i\textsuperscript{th}$ step, we define the error variable as $\varphi_{i} = x_{i} - \eta_{i - 1}\, \forall \,i=2,\cdots,n-1$, which on differentiating with respect to time and using \eqref{eq:xi_dot} yields,
\begin{equation} \label{eq:varphi_i_dot}
\dot{\varphi}_{i} = \dot{x}_{i} - \dot{\eta}_{i - 1} = f_{i}\left(\bar{x}_{i}\right) + g_{i}\left(\bar{x}_{i}\right) x_{i+1} - \dot{\eta}_{i - 1}.
\end{equation}
By considering $x_{i + 1}$ a virtual input to the $i\textsuperscript{th}$ subsystem, we choose a Lyapunov function candidate as $\mathcal{V}_{i} = \mathcal{V}_{i-1} + 0.5 \varphi_{i}^2$. By differentiating $\mathcal{V}_{i}$ with respect to time along the state trajectories, one has
\begin{equation*}
\dot{\mathcal{V}}_{i} = \dot{\mathcal{V}}_{i-1} + \varphi_{i} \dot{\varphi}_{i},
\end{equation*}
which on using \eqref{eq:varphi_i_dot}, yields
\begin{equation} \label{eq:v_i_dot_1}
 \dot{\mathcal{V}}_{i} = \dot{\mathcal{V}}_{i-1} + \varphi_{i} \left[ f_{i}\left(\bar{x}_{i}\right) + g_{i}\left(\bar{x}_{i}\right) x_{i+1} - \dot{\eta}_{i - 1} \right],   
\end{equation}
Now define $\varphi_{i+1}\coloneqq x_{i+1} - \eta_{i}$, where $\eta_{i}$ is a stabilizing function. By substituting for $x_{i+1}$ into \eqref{eq:v_i_dot_1}, one may obtain
 \begin{align} 
 \dot{\mathcal{V}}_{i} =&~ \varphi_{i} \left[ f_{i}\left(\bar{x}_{i}\right) + g_{i}\left(\bar{x}_{i}\right)\left( \varphi_{i+1} +\eta_{i} \right) - \dot{\eta}_{i - 1} \right]   \nonumber\\
 &~-\sum_{j=1}^{i-1} k_{j} \varphi_{j}^2+ g_{i-1} \left( x_{i-1}\right) \varphi_{i} .\label{eq:v_i_dot_2} 
\end{align}
The stabilizing function $\eta_{i}$ is designed as
\begin{equation}\label{eq:eta_i}
\eta_{i} = \dfrac{1}{g_{i}\left(\bar{x}_{i}\right)} \left[ \dot{\eta}_{i - 1} - f_{i}\left(\bar{x}_{i}\right) - g_{i-1}\left( x_{i-1} \right)- k_{i} \varphi_{i}\right],
\end{equation}
where $k_{i}$ is a positive design constant. On substituting the value of $\eta_{i}$ from \eqref{eq:eta_i} into \eqref{eq:v_i_dot_2}, we get
\begin{equation} \label{eq:v_i_dot_final}
 \dot{\mathcal{V}}_{i} = -\sum_{j=1}^{n-1} k_{j} \varphi_{j}^2 + g_{i}\left(\bar{x}_{i}\right) \varphi_{i+1}.
\end{equation}

\textit{Step n}: In this step, the error variable is defined as $\varphi_{n} \coloneqq x_{n} - \eta_{n-1}$, where $\eta_{n-1}$ will be obtained from the $(n-1)\textsuperscript{th}$ step. Consider a Lyapunov function candidate as $\mathcal{V}_{n} = \mathcal{V}_{n-1} +0.5 \varphi_{n}^2$, whose time derivative is obtained as $\dot{\mathcal{V}}_{n} = \dot{\mathcal{V}}_{n-1} +\varphi_{n} \dot{\varphi_{n}}$. Using \eqref{eq:v_i_dot_final}, one may write $\dot{\mathcal{V}}_{n}$ as
\begin{equation} \label{eq:v_n_dot_1}
\dot{\mathcal{V}}_{n} = -\sum_{j=1}^{n-1} k_{j} \varphi_{j}^2 + g_{i}\left(\bar{x}_{i}\right) \varphi_{i+1} + \varphi_{n} \dot{\varphi}_{n}.
\end{equation}
The derivative of $\varphi_{n}$ can be obtained using \eqref{eq:xn_dot} as
\begin{equation} \label{eq:varphi_n_dot_1}
\dot{\varphi_{n}} = \dot{x}_{n} - \dot{\eta}_{n-1} = f_{n}\left(\bar{x}_{n}\right) + g_{n}\left(\bar{x}_{n}\right) u - \dot{\eta}_{n-1}.
\end{equation}
Now, define $\varrho \coloneqq u - \eta_{n}$, where $\eta_{n}$ is a stabilizing function. Using the relation $u = \varrho + \eta_{n}$ and \eqref{eq:varphi_n_dot_1}, one may write \eqref{eq:v_n_dot_1} as
\begin{align}
\dot{\mathcal{V}}_{n} =&~ -\sum_{j=1}^{n-1} k_{j} \varphi_{j}^2 + g_{i}\left(\bar{x}_{i}\right) \varphi_{i+1} + \varphi_{n} \left[f_{n}\left(\bar{x}_{n}\right) + g_{n}\left(\bar{x}_{n}\right) u \right. \nonumber\\
&~\left.- \dot{\eta}_{n-1} \right]   \nonumber\\
=& -\sum_{j=1}^{n-1} k_{j} \varphi_{j}^2 + g_{i}\left(\bar{x}_{i}\right) \varphi_{i+1} + \varphi_{n} \left[f_{n}\left(\bar{x}_{n}\right) \right. \nonumber\\
&~\left.  + g_{n}\left(\bar{x}_{n}\right) \left( \varrho + \eta_{n} \right)- \dot{\eta}_{n-1} \right]. \label{eq:v_n_dot_2}
\end{align}
We design the stabilizing function $\eta_{n}$ as
\begin{equation}\label{eq:eta_n}
   \eta_{n} = \dfrac{ \dot{\eta}_{n-1}- f_{n}\left(\bar{x}_{n}\right) -g_{n-1}\left(\bar{x}_{n-1}\right) - k_{n} \varphi_{n} }{g_{n}\left(\bar{x}_{n}\right)}.
\end{equation}
By substituting the value of $\eta_{n}$ from \eqref{eq:eta_n} into \eqref{eq:v_n_dot_2} and performing some algebraic simplifications lead us to 
\begin{align}
\dot{\mathcal{V}}_{n}&= -\sum_{j=1}^{n-1} k_{j} \varphi_{j}^2 + g_{i}\left(\bar{x}_{i}\right) \varphi_{i+1} + \varphi_{n} \left[f_{n}\left(\bar{x}_{n}\right) + g_{n}\left(\bar{x}_{n}\right) \left\{ \varrho \right. \right. \nonumber\\
&\left.\left.  + \dfrac{1}{g_{n}\left(\bar{x}_{n}\right)} \left( \dot{\eta}_{n-1}- f_{n}\left(\bar{x}_{n}\right) -g_{n-1}\left(\bar{x}_{n-1}\right) - k_{n} \varphi_{n} \right) \right\} \right. \nonumber\\
&\left.- \dot{\eta}_{n-1} \right], \nonumber \\
&=  -\sum_{j=1}^{n} k_{j} \varphi_{j}^2 + \varphi_{n}g_{n}\left(\bar{x}_{n}\right) \varrho. \label{eq:v_n_dot_final}
\end{align}

\textit{Step n+1}: In this step, we design the commanded control input $u_{c}$. Since we have $\varrho \coloneqq u - \eta_{n}$, its time derivative can be obtained as $\dot{\varrho} = \dot{u} - \dot{\eta}_{n}$, which on substituting for $\dot{u}$ from \eqref{eq:u_saturation_model} results in 
\begin{align}
\dot{\varrho} =&~ p_1 \left[\rho  \left\{ 1 - \left(\dfrac{u}{u_{\max}}\right)^{\gamma} \right\}  + \left(1 - \rho \right)  \left\{ 1 - \left(\dfrac{u}{u_{\min}}\right)^{\gamma} \right\}  \right] u_{c} \nonumber\\
&~-p_1 p_2 u - \dot{\eta}_{n}. \label{eq:dot_varrho}
\end{align}
Consider a Lyapunov function candidate as $\mathcal{V}_{n+1} = \mathcal{V}_{n} +0.5 \varrho^2$, whose time derivative along the state trajectory is given by $\dot{\mathcal{V}}_{n+1} = \dot{\mathcal{V}}_{n} + \varrho \dot{\varrho}$, which on using the relation in \eqref{eq:dot_varrho} and \eqref{eq:v_n_dot_final} yields
\begin{align}
\dot{\mathcal{V}}_{n+1} &= -\sum_{j=1}^{n} k_{j} \varphi_{j}^2 + \varphi_{n}g_{n}\left(\bar{x}_{n}\right) \varrho + \varrho \left( \dot{u} - \dot{\eta}_{n} \right), \nonumber \\
&=-\sum_{j=1}^{n} k_{j} \varphi_{j}^2 + \varphi_{n}g_{n}\left(\bar{x}_{n}\right) \varrho + \varrho \left( p_1 \left[\rho  \left\{ 1 - \left(\dfrac{u}{u_{\max}}\right)^{\gamma} \right\} \right.\right. \nonumber\\
&~\left.\left.+ \left(1 - \rho \right)  \left\{ 1 - \left(\dfrac{u}{u_{\min}}\right)^{\gamma} \right\}  \right] u_{c} -p_1 p_2 u - \dot{\eta}_{n} \right). \label{eq:v_n+1_dot_1}
\end{align}
Now, we design the commanded control input $u_{c}$ as
\begin{equation} \label{eq:u_c}
u_{c} = \dfrac{p_1 p_2 u + \dot{\eta}_{n} - \varphi_{n}g_{n}\left(\bar{x}_{n}\right) -k_{n+1}\varrho}{p_1 \left[\rho  \left\{ 1 - \left(\dfrac{u}{u_{\max}}\right)^{\gamma} \right\}  + \left(1 - \rho \right)  \left\{ 1 - \left(\dfrac{u}{u_{\min}}\right)^{\gamma} \right\}  \right]} .
\end{equation}
With the commanded control $u_{c}$ given in \eqref{eq:u_c}, the derivative of Lyapunov function candidate given in \eqref{eq:v_n+1_dot_1} becomes
\begin{align*}
\dot{\mathcal{V}}_{n+1} =&~-\sum_{j=1}^{n} k_{j} \varphi_{j}^2 + \varphi_{n}g_{n}\left(\bar{x}_{n}\right) \varrho + \varrho \left[ -p_1 p_2 u - \dot{\eta}_{n}   \right.\\
&~\left.+ p_1 \left[\rho  \left\{ 1 - \left(\dfrac{u}{u_{\max}}\right)^{\gamma} \right\}  + \left(1 - \rho \right)  \left\{ 1 - \left(\dfrac{u}{u_{\min}}\right)^{\gamma} \right\}  \right]\times \right. \nonumber\\
&~\left. \dfrac{p_1 p_2 u + \dot{\eta}_{n} - \varphi_{n}g_{n}\left(\bar{x}_{n}\right) -k_{n+1}\varrho}{p_1 \left[\rho  \left\{ 1 - \left(\dfrac{u}{u_{\max}}\right)^{\gamma} \right\}  + \left(1 - \rho \right)  \left\{ 1 - \left(\dfrac{u}{u_{\min}}\right)^{\gamma} \right\}  \right]} \right]
\end{align*}
which, after some algebraic simplifications, results in
\begin{equation} \label{eq:v_n+1_dot_final}
\dot{\mathcal{V}}_{n+1} = -\sum_{j=1}^{n} k_{j} \varphi_{j}^2 - k_{n+1}\varrho^2,
\end{equation}
which is negative semi-definite. Thus, the system is considered stable in Lyapunov's sense.  In the following theorem, we prove a stronger notion of stability of the closed-loop system.
\begin{theorem}\label{thm:global_tracking}
Consider the closed-loop system consisting of nonlinear plant \eqref{eq:system_dynamics} and the input saturation model \eqref{eq:u_saturation_model}. Under the virtual control inputs \eqref{eq:eta_1}, \eqref{eq:eta_i}, \eqref{eq:eta_n} and commanded input \eqref{eq:u_c}, the closed-loop system is globally asymptotically stable.
\end{theorem}
\begin{proof}
Consider a radially unbounded Lyapunov function candidate as $\mathcal{V}_{n+1}(\varphi_{1:n}(t),\varrho(t))$, whose time derivative along the trajectories of the system can be obtained using \eqref{eq:v_n+1_dot_final} as $\dot{\mathcal{V}}_{n+1}(\varphi_{1:n}(t),\varrho(t)) = -\sum_{j=1}^{n} k_{j} \varphi_{j}(t)^2 - k_{n+1}\varrho(t)^2 \leq 0$, which is globally negative definite on $\mathbb{R}^{n+1}$ for all $\varphi_{i}(t)$ ($i=1,2,\cdots,n$) and $\varrho$ not equal to zero. Thus, $\dot{\mathcal{V}}_{n+1}(\varphi_{1:n}(t),\varrho(t))$ is monotonically deceasing function of $t$. Note that the Lyapunov function candidate is a continuously differentiable positive definite function on $\mathbb{R}^{n+1}$. Now, let $\mathbb{E}$ be a set defined as $\mathbb{E} \coloneqq \{ \mathbf{X}(t) \in \mathbb{R}^{n+1} \rvert \dot{\mathcal{V}}_{n+1}(\mathbf{X}(t))=0\} $, where $\mathbf{X}(t) \coloneqq[\varphi_{1:n},\varrho]^\top $. Clearly, the only solution $\mathbf{X}(t)$ that can stay in $\mathbb{E}$ is the trivial solution $\mathbf{X}(t) \equiv 0$. Hence, using the Lyapunov stability theory, we conclude that $[\varphi_{1:n}(t),\varrho(t)]^\top \to 0$ as $t \to \infty$ $\forall$ $[\varphi_{1:n}(0),\varrho(0)]^\top \in \mathbb{R}^{n+1}$.
\end{proof}

While Theorem~\ref{thm:global_tracking} ensures that the closed-loop system is globally asymptotically stable, the following theorem establishes that all closed-loop signals, including the commanded control input  $u_c$, remain bounded. Once we show that $u_c$ is bounded, then by virtue of Theorem~\ref{thm:sat_model}, it is guaranteed that the plant input $u$ will remain confined in the set $\mathbb{U}$, as the boundedness of $u_c$ is a sufficient condition for Theorem~\ref{thm:sat_model}.
\begin{theorem} \label{thm:boundedness}
Consider the nonlinear closed-loop system governed by the plant dynamics in \eqref{eq:system_dynamics} and the input saturation model as in \eqref{eq:u_saturation_model}. Suppose that the stabilizing functions are designed as \eqref{eq:eta_1}, \eqref{eq:eta_i}, \eqref{eq:eta_n}, and the commanded control input given by \eqref{eq:u_c}, then, for any initial condition, all closed-loop signals are uniformly bounded, and the plant input remains within the set $\mathbb{U}$.
\end{theorem}

\begin{proof}
We first establish the boundedness of closed-loop signals using mathematical induction. From \eqref{eq:v_n+1_dot_final}, we have $\dot{\mathcal{V}}_{n+1} = -\sum_{j=1}^{n} k_{j} \varphi_{j}^2 - k_{n+1}\varrho^2$, which one may write as $\dot{\mathcal{V}}_{n+1} \leq - \vartheta_{1} \mathcal{V}_{n+1}$, where $\vartheta_{1} \coloneqq \min\{2k_{1},2k_{2},\cdots,2k_{n+1}\}$ is a strictly positive constant. This, on integrating within  suitable limits, yields $\mathcal{V}_{n+1}(t) \leq  \mathcal{V}_{n+1}(0)e^{- \vartheta_{1} t}$, which on expanding can be written as $0.5\sum_{j=1}^{n} \varphi_{j}(t)^2+ 0.5\varrho(t)^{2} \leq  \mathcal{V}_{n+1}(0)e^{- \vartheta_{1} t}$. This implies that $\|\mathbf{X}(t)\| \leq  \mathcal{V}_{n+1}(0)e^{- \vartheta_{1} t}$, where $\mathbf{X}(t) \coloneqq [\varphi_{1:n}(t), \varrho(t)]^{\top}$. Thus, the signals $\varphi_{j}(t)$ for $j=1,2,\cdots,n$ and $\varrho$ are bounded. From the relation $\varphi_1 = y - y_d = x_{1} - y_{d}$ and given that $y_{d}$ is bounded it follows that $y$ and $x_{1}$ are also bounded since $y=x_{1}$. It can be noted from the right-hand side of \eqref{eq:eta_1} that the stabilizing function $\eta_1$ is a function of bounded signals $x_{1}$, $\dot{y}_{d}$, ensuring $\eta_{1}$ is bounded. The boundedness of $\eta_{1}$ together with the boundedness of $\varphi_{1}$ implies that $x_{2} = \varphi_{2} + \eta_{1}$ remains bounded. By induction, we can prove that the stabilizing functions $\eta_{i}$ for $i=2,3,\cdots,n$ remain bounded. Consequently, the boundedness of the states $x_{i}$ can be proved using the boundedness of $\eta_{i-1}$ and $\varphi_{i}$. Finally, the boundedness of $\eta_{n}$ and $\varrho$ implies that the state $u$ is bounded. Since $u_{c}$  in \eqref{eq:u_c}  is expressed as a function of $\varrho$ and states $\bar{x}_{n}$ and $u$, which are bounded, it follows that $u_{c}$ is bounded. This, in turn, together with Theorem~\ref{thm:sat_model} guarantees the plant input remains confined to the set $\mathbb{U}$.
\end{proof}
\subsection{Constrained Output Tracking with Bounded Control}
In many practical applications, it is often necessary to impose additional constraints on the system's output to ensure safe and reliable operation. Such constraints may arise from operational limits, safety considerations, or mission-specific requirements.  To address this, in this subsection, we extend our proposed input saturation model to incorporate output constraints. Thus, in the following subsection, we present a framework that seamlessly integrates the input saturation model with time-varying output constraints, ensuring that the nonlinear plant output and input both remain within a predefined bound, that is, $u$ remains confined to the set $\mathbb{U}$, whereas $y(t)$ remains confined to the set $\mathbb{Y}$. Here, $\mathbb{Y}$ is defined as $\mathbb{Y} \coloneq \{ y(t) \in \mathbb{R}\, \rvert\, \underline{y}(t) < y(t) <\overline{y}(t)\}$ for all time $t \geq 0$ with $\underline{y}(t)$ and $\overline{y}(t)$ being the time varying constraints given as $\underline{y}(t):\mathbb{R}_{+} \to \mathbb{R}$ and $\overline{y}(t):\mathbb{R}_{+} \to \mathbb{R}$ such that $\overline{y}(t) > \underline{y}(t)\, \forall\, t \geq 0$. Before proceeding with the design of the control strategy, we first present certain frequently used assumptions in the case of time-varying output constraint problems.
\begin{assumption} \label{y_bound_asy}
There exist constants $\overline{\psi}_{i}$, $\underline{\psi}_{i}$ for $i=0,1,\ldots,n+1$, such that $\underline{y}(t) \geq \underline{\psi}_{0}$, $\overline{y}(t) \leq \overline{\psi}_{0}$, $\lvert\underline{y}^{(i)}(t)\rvert \leq \underline{\psi}_{i}$, and $\lvert \overline{y}^{(i)}(t)\rvert \leq \overline{\psi}_{i}$, for $i=1,2,\ldots,n+1$ and $t \geq 0$. 
\end{assumption}
\begin{assumption} \label{yd_bound_asy}
There exist functions $\underline{\Upsilon}:\mathbb{R}_{+}\to \mathbb{R}$ and $\overline{\Upsilon}:\mathbb{R}_{+}\to \mathbb{R}$ satisfying $\underline{\Upsilon}(t) > \underline{y}(t)$ and $\overline{\Upsilon}(t) < \overline{y}(t)$ for all time $t \geq 0$, and positive constants $\mu_{i}$ for $i=1$, $2$, $\ldots$, $n+1$ such that the desired trajectories and their time derivatives satisfy $-\underline{\Upsilon}(t) \leq y_d(t) \leq \overline{\Upsilon}(t)$, $\lvert {y}^{(i)}_d(t) \rvert < \mu_i$ for all time $t \geq 0$, where $(\cdot)^{(i)}$ denotes the $i\textsuperscript{th}$ derivative of function $(\cdot)$.
\end{assumption}
Assumption \ref{yd_bound_asy} essentially means that the desired trajectory is well-behaved over time, that is, the desired trajectory and its derivative up to ($n+1$)\textsuperscript{th} order are bounded.

While various methods have been proposed to handle the output constraints, in this work, we employ the concept of barrier Lyapunov function proposed in \cite{doi:10.1016/j.automatica.2008.11.017}. We consider that the output of the plant is subjected to a time-varying constraint. The primary motivation for considering time-varying output constraints lies in the fact that the symmetric and constant bounds can always be considered special cases. Thus, this consideration will give a generic treatment to the output constraint problems.

We again consider the augmented nonlinear system given in \eqref{eq:system_dynamics_aug_global}. Following our previous design, we again leverage the concept of \emph{backstepping} control to design the commanded control input $u_{c}$ in this case. For conciseness, we provide detailed derivations only for the novel aspects of the control design that differ substantially from the previous case, while summarizing analogous steps in compact form to maintain focus on the key technical contributions.

\textit{Step 1}: We employ the same coordinate transformation previously used for the unconstrained output tracking problem. Consequently, the first transformed state $\varphi_1$ is again defined as the tracking error $\varphi_{1}\coloneq y - y_d$ maintaining consistency with the unconstrained case. 
To restrict the output within a specified set, we consider a barrier Lyapunov function candidate as
\begin{align} 
\mathcal{W}_{1} =&~ \dfrac{s\left( \varphi_{1} \right)}{2r} \ln\left( \dfrac{\beta\left( t \right)^{2r}}{\beta\left( t \right)^{2r}-\varphi_{1}^{2r}} \right) + \dfrac{ 1 - s\left( \varphi_{1} \right) }{2r}  \nonumber\\&~\times\ln\left( \dfrac{\alpha\left( t \right)^{2r}}{\alpha\left( t \right)^{2r}-\varphi_{1}^{2r}} \right) \label{eq:w},
\end{align}
where $r \in \mathbb{N}$ such that $2r \geq n$, the function $s(\cdot)$ is defined as $s(\cdot)=1$ if $(\cdot)>0$ and $s(\bullet)=0$ otherwise, and $\alpha(t)$ and $\beta(t)$ are time-varying lower and upper bounds of $\varphi_1(t)$, which are defined as $\alpha(t) \coloneqq y_d(t) - \underline{y}(t)$ and $\beta(t) \coloneqq \overline{y}(t) -  y_d(t)$. From hereafter, we drop the argument of $s$ and $t$ for brevity and to streamline the discussion. 

Using Assumptions \ref{assum:1} and \ref{y_bound_asy}, there exist positive constants $\underline{\alpha}$, $\overline{\alpha}$ $\underline{\beta}$, and $\overline{\beta}$, that satisfy $\underline{\alpha} \leq \alpha(t)\leq \overline{\alpha}$ and $\underline{\beta} \leq \beta(t)\leq \overline{\beta}$ for all time $t \geq 0$. The Lyapunov function candidate given in \eqref{eq:w} can be written as $\mathcal{W}_{1} = \frac{1}{2r} \ln\left( \frac{1}{1 - \zeta^{2r}} \right)$,
by performing a change of coordinates as $\underline{\zeta}=\varphi_{1}/\alpha$ and $\overline{\zeta}=\varphi_{1}/\beta$, and choosing $\zeta = s \overline{\zeta} + \left( 1 -s \right) \underline{\zeta}$. It readily follows that the $\mathcal{W}_{1}$ is positive definite and continuously differentiable in the set $|\zeta| < 1$, and thus it is a valid Lyapunov candidate. On differentiating $\mathcal{W}_{1}$ with respect to time along the state trajectories yields
\begin{equation*}
\dot{\mathcal{W}}_{1} = \dfrac{\left( 1 - \zeta^{2r}\right)}{2r}  \left[ \dfrac{1}{\left( 1 - \zeta^{2r}\right)^{2}} \right] \left(- 2r \zeta^{2r-1} \dot{\zeta} \right) = \dfrac{\zeta^{2r-1}}{\left( 1 - \zeta^{2r}\right)} \dot{\zeta},
\end{equation*}
which on substituting the value of $\dot{\zeta}$ using $\zeta = s \underline{\zeta} + \left( 1 -s \right) \overline{\zeta}$, \eqref{eq:varphi1_dot}, and the relation $x_{2}=\varphi_2 + \eta_{1}$ lead us to arrive at
\begin{align}
&\dot{\mathcal{W}}_{1} = \dfrac{s\overline{\zeta}^{2r-1}}{\alpha\left( 1 - \overline{\zeta}^{2r}\right)} \left( f_{1}(x_1) + g_1(x_{1}) x _{2}- \dot{y}_{d}  - \varphi_{1} \dfrac{\dot{\alpha}}{\alpha}\right) \nonumber\\
& +\dfrac{(1-s)\underline{\zeta}^{2r-1}}{\beta\left( 1 - \underline{\zeta}^{2r}\right)} \left[ f_{1}(x_1) + g_1(x_{1}) x _{2}- \dot{y}_{d}  - \varphi_{1} \dfrac{\dot{\beta}}{\beta}\right], \nonumber\\
&= \dfrac{s\overline{\zeta}^{2r-1}}{\alpha\left( 1 - \overline{\zeta}^{2r}\right)} \left( f_{1}(x_1) + g_1(x_{1}) \left(\varphi_2 + \eta_{1}\right)- \dot{y}_{d}  - \varphi_{1} \dfrac{\dot{\alpha}}{\alpha}\right) \nonumber\\
& +\dfrac{(1-s)\underline{\zeta}^{2r-1}}{\beta\left( 1 - \underline{\zeta}^{2r}\right)} \left[ f_{1}(x_1) + g_1(x_{1}) \left(\varphi_2 + \eta_{1}\right)- \dot{y}_{d}  - \varphi_{1} \dfrac{\dot{\beta}}{\beta}\right]\label{eq:w_1_dot}.
\end{align}
Here we design the stabilizing function $\eta_{1}$ as
\begin{equation}\label{eq:eta_1_asy}
\eta_{1} = \dfrac{1}{g_{1}\left( x_{1} \right)}\left[ \dot{y}_{d} - f_{1}(x_1) - \left( k_{1} + \overline{k}_{1}(t) \right) \varphi_{1} \right],
\end{equation}
where $k_{1}$ is a positive constant and $\overline{k}_{1}(t)$ is a time-varying gain chosen as
\begin{equation} \label{eq:k_1_t}
\overline{k}_{1}(t) = \sqrt{\left( \dfrac{\dot{\overline{\zeta}}}{\overline{\zeta}} \right)^2+\left( \dfrac{\dot{\underline{\zeta}}}{\underline{\zeta}} \right)^{2}+\delta}
\end{equation}
with a positive constant $\delta$. After substituting the value of $\eta_{1}$ from \eqref{eq:eta_1_asy} and noticing that $\overline{k}+\frac{s \dot{\beta}}{\beta} + \frac{\left(1-s\right)\dot{\alpha}}{\alpha} \geq 0$, one may obtain the derivative $\dot{\mathcal{W}}_{1}$ given in \eqref{eq:w_1_dot} as
\begin{equation}\label{eq:w_1_dot_final}
\dot{\mathcal{W}}_{1} =-\dfrac{k_{1}\zeta^{2r}}{1-\zeta^{2r}} + \digamma  g_1(x_{1}) \varphi^{2r-1} \varphi_{2},
\end{equation}
where $\digamma$ is given as $\digamma = \frac{s}{\beta^{2r}-\varphi_{1}^{2r}} +\frac{1-s}{\alpha^{2r}-\varphi_{1}^{2r}}$.
\begin{remark}
The constant $\beta$ in \eqref{eq:k_1_t} ensures boundedness of $\dot{\eta}_{1}$ even if $\dot{\overline{\zeta}}$ and $\dot{\underline{\zeta}}$ are zero.    
\end{remark}

\textit{Step i ($2 \leq i \leq n-1$)}: Similar to the unconstrained output case, we define the error variable for the $i\textsuperscript{th}$ step as $\varphi_{i}$, whose derivative is given in \eqref{eq:varphi_i_dot}. We choose a Lyapunov function candidate as $\mathcal{W}_{i} = \mathcal{W}_{i-1} + 0.5 \varphi_{i}^2$, whose time derivative along the state trajectories is given by $\dot{\mathcal{W}}_{i} = \dot{\mathcal{W}}_{i-1} + \varphi_{i} \dot{\varphi}_{i}$,
which on using \eqref{eq:w_1_dot_final} and \eqref{eq:varphi_i_dot}, yields
\begin{align*} 
 \dot{\mathcal{W}}_{i} =&~ -\dfrac{k_{1}\zeta^{2r}}{1-\zeta^{2r}} + \digamma  g_1(x_{1}) \varphi^{2r-1} \varphi_{2} + \varphi_{i} \left[ f_{i}\left(\bar{x}_{i}\right) \right.\nonumber \\
 &~\left. + g_{i}\left(\bar{x}_{i}\right) x_{i+1} -\dot{\eta}_{i - 1} \right],   
\end{align*}
which on using the relation $x_{i+1} = \varphi_{i+1}+\eta_{i}$, and  substituting for $x_{i+1}$ results in 
\begin{align} 
 \dot{\mathcal{W}}_{i} =&~ -\dfrac{k_{1}\zeta^{2r}}{1-\zeta^{2r}} + \digamma  g_1(x_{1}) \varphi^{2r-1} \varphi_{2} + \varphi_{i} \left( f_{i}\left(\bar{x}_{i}\right) \right.\nonumber \\
 &~\left.+ g_{i}\left(\bar{x}_{i}\right) \left(  \varphi_{i+1}+\eta_{i} \right) - \dot{\eta}_{i - 1} \right).   \label{eq:w_i_dot_1}
\end{align}
The stabilizing function $\eta_{i}$ is designed as
\begin{equation}\label{eq:eta_i_asy}
\eta_{i} = \dfrac{ \dot{\eta}_{i - 1} - f_{i}\left(\bar{x}_{i}\right) - \digamma  g_1(x_{1}) \varphi^{2r-1} - k_{i} \varphi_{i}}{g_{i}\left(\bar{x}_{i}\right)} .
\end{equation}
On substituting for $\eta_{i}$ from \eqref{eq:eta_i_asy} into \eqref{eq:w_i_dot_1}, one may have
\begin{align} 
 \dot{\mathcal{W}}_{i} =&~ -\dfrac{k_{1}\zeta^{2r}}{1-\zeta^{2r}} - \sum_{j=2}^{n-1} k_{j} \varphi_{j}^2 + g_{i}\left(\bar{x}_{i}\right) \varphi_{i+1} . \label{eq:w_i_dot_final}
\end{align}

\textit{Step n}: The error variable in this step is defined as $\varphi_n \coloneqq x_{n} - \eta_{n-1}$, where $\eta_{n-1}$ will be obtained from the $(n-1)\textsuperscript{th}$ step. The derivative of $\varphi_{n}$ is given by \eqref{eq:varphi_i_dot}. Now, consider a Lyapunov function candidate as $\mathcal{W}_{n} = \mathcal{W}_{n-1} +0.5 \varphi_{n}^2$, whose time derivative is given as $\dot{\mathcal{W}}_{n} = \dot{\mathcal{W}}_{n-1} +\varphi_{n} \dot{\varphi_{n}}$. Using \eqref{eq:w_i_dot_final}, one may write $\dot{\mathcal{W}}_{n}$ as
\begin{equation} \label{eq:W_n_dot_1}
\dot{\mathcal{W}}_{n} =-\dfrac{k_{1}\zeta^{2r}}{1-\zeta^{2r}} - \sum_{j=2}^{n-1} k_{j} \varphi_{j}^2 + g_{i}\left(\bar{x}_{i}\right) \varphi_{i+1} + \varphi_{n} \dot{\varphi}_{n}.
\end{equation}
Similar to the unconstrained output case, we define $\varrho \coloneqq u - \eta_{n}$, where $\eta_{n}$ is a stabilizing function. Now, using the relation $u = \varrho + \eta_{n}$ and \eqref{eq:varphi_n_dot_1}, one may write \eqref{eq:W_n_dot_1} as
\begin{align}
&\dot{\mathcal{W}}_{n} = -\dfrac{k_{1}\zeta^{2r}}{1-\zeta^{2r}} - \sum_{j=2}^{n-1} k_{j} \varphi_{j}^2 + g_{i}\left(\bar{x}_{i}\right) \varphi_{i+1}+ \varphi_{n} \left(f_{n}\left(\bar{x}_{n}\right)\right. \nonumber\\
&\left.  + g_{n}\left(\bar{x}_{n}\right) u  - \dot{\eta}_{n-1} \right)  = -\dfrac{k_{1}\zeta^{2r}}{1-\zeta^{2r}} - \sum_{j=2}^{n-1} k_{j} \varphi_{j}^2 + g_{i}\left(\bar{x}_{i}\right) \varphi_{i+1}\nonumber\\
&   + \varphi_{n} \left(f_{n}\left(\bar{x}_{n}\right)  + g_{n}\left(\bar{x}_{n}\right) \left( \varrho + \eta_{n} \right)  - \dot{\eta}_{n-1} \right) \nonumber,
\end{align}
which on choosing the stabilizing function $\eta_{n}$ the same as in \eqref{eq:eta_n}
and performing some algebraic simplifications yields
\begin{align}
\dot{\mathcal{W}}_{n}&= -\dfrac{k_{1}\zeta^{2r}}{1-\zeta^{2r}} - \sum_{j=2}^{n-1} k_{j} \varphi_{j}^2 + g_{i}\left(\bar{x}_{i}\right) \varphi_{i+1}+ \varphi_{n} \left[f_{n}\left(\bar{x}_{n}\right)  \right. \nonumber\\
&\left. + g_{n}\left(\bar{x}_{n}\right) \left( \varrho  + \dfrac{1}{g_{n}\left(\bar{x}_{n}\right)} \left( \dot{\eta}_{n-1}- f_{n}\left(\bar{x}_{n}\right) -g_{n-1}\left(\bar{x}_{n-1}\right) \right. \right.\right. \nonumber\\
&\left.\left.\left.- k_{n} \varphi_{n} \right) \right)- \dot{\eta}_{n-1} \right] \nonumber\\
&= -\dfrac{k_{1}\zeta^{2r}}{1-\zeta^{2r}} -\sum_{j=2}^{n} k_{j} \varphi_{j}^2 + \varphi_{n}g_{n}\left(\bar{x}_{n}\right) \varrho. \label{eq:w_n_dot_final}
\end{align}

\textit{Step n+1:}  Consider a Lyapunov function candidate as $\mathcal{W}_{n+1} = \mathcal{W}_{n} +0.5 \varrho^2$, whose time derivative along the state trajectory is obtained as $\dot{\mathcal{W}}_{n+1} = \dot{\mathcal{W}}_{n} + \varrho \dot{\varrho}$. Upon using the relation $\dot{\varrho} = \dot{u} - \dot{\eta}_{n}$ and substituting for $\dot{u}$ from \eqref{eq:u_saturation_model} and using the expression \eqref{eq:w_n_dot_final} results in
\begin{align}
\dot{\mathcal{W}}_{n+1}& = -\dfrac{k_{1}\zeta^{2r}}{1-\zeta^{2r}} -\sum_{j=2}^{n} k_{j} \varphi_{j}^2 + \varphi_{n}g_{n}\left(\bar{x}_{n}\right) \varrho + \varrho \left( \dot{u} - \dot{\eta}_{n} \right), \nonumber \\
=& -\sum_{j=2}^{n} k_{j} \varphi_{j}^2 + \varphi_{n}g_{n}\left(\bar{x}_{n}\right) \varrho + \varrho \left( g(u) u_{c} -f(u) - \dot{\eta}_{n} \right) \nonumber\\
&-\dfrac{k_{1}\zeta^{2r}}{1-\zeta^{2r}}. \label{eq:w_n+1_dot_1}
\end{align}
With the commanded control input $u_{c}$ given in \eqref{eq:u_c}, the time derivative of Lyapunov function candidate given in \eqref{eq:w_n+1_dot_1} becomes
\begin{align*}
&\dot{\mathcal{W}}_{n+1} =-\dfrac{k_{1}\zeta^{2r}}{1-\zeta^{2r}}  + \varphi_{n}g_{n}\left(\bar{x}_{n}\right) \varrho+ \varrho \left[ g(u) \left\{ \dfrac{1}{g(u)} \left(f(u) \right.\right.\right.\\
&\left.\left.\left.+\; \dot{\eta}_{n} - \varphi_{n}g_{n}\left(\bar{x}_{n}\right) -k_{n+1}\varrho \right) \right\} -f(u) - \dot{\eta}_{n} \right] -\sum_{j=2}^{n} k_{j} \varphi_{j}^2
\end{align*}
which, after some algebraic simplifications, results in
\begin{align}
\dot{\mathcal{W}}_{n+1} =&~ -\sum_{j=2}^{n} k_{j} \varphi_{j}^2 - k_{n+1}\varrho^2 -\dfrac{k_{1}\zeta^{2r}}{1-\zeta^{2r}}.\label{eq:w_n+1_dot_final}
\end{align}
\begin{theorem} \label{thm:semi_global_tracking}
Consider the closed-loop system consisting of the plant \eqref{eq:system_dynamics} and the proposed input saturation model \eqref{eq:u_saturation_model}, where Assumptions \ref{assum:1}, \ref{y_bound_asy}, and \ref{yd_bound_asy} hold true. If the initial value of the system's output satisfies $\alpha(0) < y(0) < \beta(0)$ with $u(0) = 0$, then under the stabilizing functions \eqref{eq:eta_1}, \eqref{eq:eta_i}, \eqref{eq:eta_n} and the commanded control input \eqref{eq:u_c} the closed-loop system is semi-globally asymptotically stable.
\end{theorem}
\begin{proof}
The proof is omitted here for brevity, as it follows a similar approach to that of Theorem~\ref{thm:global_tracking}.
\end{proof}
\begin{remark}
The term semi-global is due to the fact that the initial value of the output needs to be in a predefined set.
\end{remark}
\begin{theorem} \label{thm:error_bound_asy}
Consider the closed-loop system \eqref{eq:system_dynamics}, \eqref{eq:u_saturation_model}, \eqref{eq:u_c}, and let Assumptions \ref{assum:1}, \ref{y_bound_asy}, and \ref{yd_bound_asy} hold true. If the initial value of the system's output satisfies $\alpha(0) < y(0) < \beta(0)$ with $u(0) = 0$, then the following holds:
\begin{itemize}
    \item The system's output remains confined to the set $\mathbb{Y}$.
    \item All closed-loop signals are bounded.
    \item The system's input $u$ remains confined to the set $\mathbb{U}$.
\end{itemize}
\end{theorem}
\begin{proof}
Let the initial value of output satisfy the condition $\beta(0) < y(0) < \alpha(0)$. Since $\varphi_{1} = y - y_{d}$, we can write the above inequality in terms of $\varphi_{1}$ as $\alpha(0) < \varphi_{1}(0) < \beta(0)$ using the relation $\alpha(t) = y_{d}(t) - \underline{y}(t)$ and $\beta(t) =  \overline{y}(t) - y_d(t)$. This, in turn, is equivalent to the condition $\lvert \zeta(0) \rvert <1$ as $\lvert\zeta(t)\rvert < 1$ $\iff$ $\alpha(t) < \varphi_{1}(t) < \beta(t)$. Using the results in Lemma~\ref{lem:BLF}, one can conclude that $\lvert \zeta(t)\rvert < 1$ for all time $t \geq 0$. Since, $\lvert \zeta(t) \rvert < 1$, we have $-\alpha(t) < \varphi_{1}(t) < \beta(t)$. Using the relation $\varphi_{1}(t) = y(t) - y_d(t)$, one can express $y(t) = \varphi_{1}(t) + y_d(t)$, which can be written as $-\alpha(t) +y_d(t) < \varphi_{1}(t) +y_d(t) = y(t) < \beta(t)+y_d(t)$. Using the definition of $\alpha(t)$ and $\beta(t)$, we conclude that $\underline{y}(t) < y(t) < \overline{y}(t)$ for all times $t \geq 0$.

We first analytically calculate the bounds on the output tracking error $\varphi_{1}$ and then use mathematical induction to prove that all remaining closed-loop signals are bounded and $u$ remains confined to the set $\mathbb{U}$. Note that the expression in \eqref{eq:w_n+1_dot_final} can also be written as $\dot{\mathcal{W}}_{n+1} \leq - \vartheta_{2} \mathcal{W}_{n+1}$, using Lemma~\ref{eq:blf_log_inequality}, where $\vartheta_{2}>0$ is a constant, defined as $\vartheta_{2} \coloneqq \min\{2rk_{1},2k_{2},\cdots,2k_{n+1}\}$. By integrating the inequality within a suitable limit, one may obtain $\mathcal{W}_{n+1}(t) \leq \mathcal{W}_{n+1}(0)e^{-\vartheta_{2} t}$. As $\mathcal{W}_{n+1} = \sum_{j=1}^{n} \mathcal{W}_{j} + 0.5 \varrho^2$ is a sum of nonnegative functions, the inequality will hold for each component. Thus, we can write this inequality on $\mathcal{W}_{1}$ as $\frac{1}{2r} \ln\left( \frac{1}{1 - \zeta^{2r}} \right) \leq \mathcal{W}_{n+1}(0)e^{-\vartheta_{2} t} $, which leads us to arrive at $\ln \left(1 - \zeta^{2r}\right) \leq -2r \mathcal{W}_{n+1}(0)e^{-\vartheta_{2} t}$. This implies
\begin{equation} \label{eq:zeta_expression}
1 - \zeta^{2r} \leq e^{-2r \mathcal{W}_{n+1}(0)e^{-\vartheta_{2} t} } \Rightarrow \zeta^{2r} \leq 1 - e^{-2r \mathcal{W}_{n+1}(0)e^{-\vartheta_{2} t}}.
\end{equation}
Now, recall the relation $\zeta = s \overline{\zeta} + \left( 1 -s \right) \underline{\zeta}$ and the definition of $s$. So when $\varphi_1 > 0$, $s=1$ and the expression in \eqref{eq:zeta_expression} becomes $\overline{\zeta}^{2r} \leq 1 - e^{-2r \mathcal{W}_{n+1}(0)e^{\vartheta_{2} t}}$, which can be further expressed using the relation $\overline{\zeta} = \varphi_{1}/\beta$ as  $(\varphi_{1}/\beta)^{2r} \leq 1 - e^{-2r \mathcal{W}_{n+1}(0)e^{\vartheta_{2} t}}$, implying $\varphi_{1} \leq \beta \left( 1 - e^{-2r \mathcal{W}_{n+1}(0)e^{-\vartheta_{2} t}}\right) ^{1/2r}$. Similarly, when $\varphi_1 \leq 0$, $s=0$ and the expression in \eqref{eq:zeta_expression} becomes $\underline{\zeta}^{2r} \leq 1 - e^{-2r \mathcal{W}_{n+1}(0)e^{-\vartheta_{2} t}}$, which results in $\varphi_{1} \geq -\alpha \left( 1 - e^{-2r \mathcal{W}_{n+1}(0)e^{-\vartheta_{2} t}}\right) ^{1/2r}$ using the relation $\underline{\zeta} = \varphi_{1}/\alpha$. Therefore, $\varphi_{1}$ remains confined to the set $\mathbb{P}$ which is defined as $\mathbb{P} \coloneqq \{\varphi_{1} \rvert  \underline{\varphi}_{1} \leq  \varphi_{1}(t) \leq \overline{\varphi}_{1}\}$, where $\underline{\varphi}_{1} \coloneqq -\alpha \left( 1 - e^{-2r \mathcal{W}_{n+1}(0)e^{-\vartheta t}}\right) ^{1/2r}$ and $\overline{\varphi}_{1} \coloneqq \beta \left( 1 - e^{-2r \mathcal{W}_{n+1}(0)e^{-\vartheta t}}\right)^{1/2r}$. 

Now, we move forward to prove the boundedness of error signals $\varphi_{i}$ for $i=2,3,\ldots,n$, and $\varrho$. Using the fact that $\mathcal{W}_{n+1}$ can also be written as $\mathcal{W}_{n+1} = \mathcal{W}_{1} + \sum_{j=2}^{n} \mathcal{W}_{j} + 0.5 \varrho^2$, one may write $\mathcal{W}_{n+1}(t) \leq \mathcal{W}_{n+1}(0)e^{\vartheta_{2} t}$ and $0.5 \sum_{j=2}^{n} \varphi_{j}^2 + 0.5 \varrho^2 \leq \mathcal{W}_{n+1}(0)e^{\vartheta_{2} t}$, and thus $\| \varphi_{2:n} \| \leq \sqrt{2\mathcal{W}_{n+1}(0)e^{-\vartheta_{2} t}}$ and $\lvert \varrho \rvert \leq \sqrt{\mathcal{W}_{n+1}(0)e^{-\vartheta_{2} t}}$. Therefore, the error signals $\varphi_{i}$ for $i=1,2,\ldots,n$ and $\varrho$ are bounded as we have already proved the boundedness of $\varphi_{1}$. Recall, $y=x_{1}$ and $\varphi_{1} = y - y_{d}$. The boundedness of $\varphi_{1}$ together with Assumption~\ref{y_bound_asy} implies that $x_{1}$ is bounded. Using relations, $\underline{\alpha} \leq \alpha(t)\leq \overline{\alpha}$ and $\underline{\beta} \leq \beta(t)\leq \overline{\beta}$, the time-varying bounds can be transformed into constant bounds as $-\underline{\alpha} < \varphi_{1} < \overline{\beta}$. We can further estimate the bounds on $\lvert \dot{\beta} \rvert$ and $\lvert \dot{\alpha} \rvert$ as $\lvert \dot{\alpha} \rvert \leq \mu_{1} + \underline{y} $ and $\lvert \dot{\beta} \rvert\leq \mu_{1} + \overline{y}$  inline with Assumptions~\ref{y_bound_asy} and \ref{yd_bound_asy}, which implies $\overline{k}_{1}(t)$ in \eqref{eq:k_1_t} is bounded. 

Now, one can notice from the right-hand side of \eqref{eq:eta_1_asy} that the stabilizing function $\eta_{1}$ is a function of bounded signals. Thus, $\eta_{1}$ is bounded. This, in turn, together with the boundedness of $\varphi_{2}$ implies $x_{2}$ is bounded. By induction, we can show the $\eta_{i}$ for $i=2,3,\ldots,n$, are bounded. Consequently, the boundedness of states $x_{i}$ can be shown using the boundedness of $\eta_{i-1}$ and $\varphi_{i}$. Since the variables $\varrho$ and $\eta_{n}$ are bounded together, this implies that state $u$ is bounded. Thus, we can conclude from \eqref{eq:u_c} that $u_c$ is bounded as it is a function of bounded signals. Once $u_c$ is finite, from Theorem~\ref{thm:sat_model}, we can conclude that $u$ remains confined in the safe set $\mathbb{U}$. This concludes the proof.
\end{proof}

\section{Numerical Simulations} \label{sec:simulations}
\begin{figure*}[!ht]
\centering
\begin{subfigure}{0.25\linewidth}
	\centering
    \includegraphics[width=\linewidth]{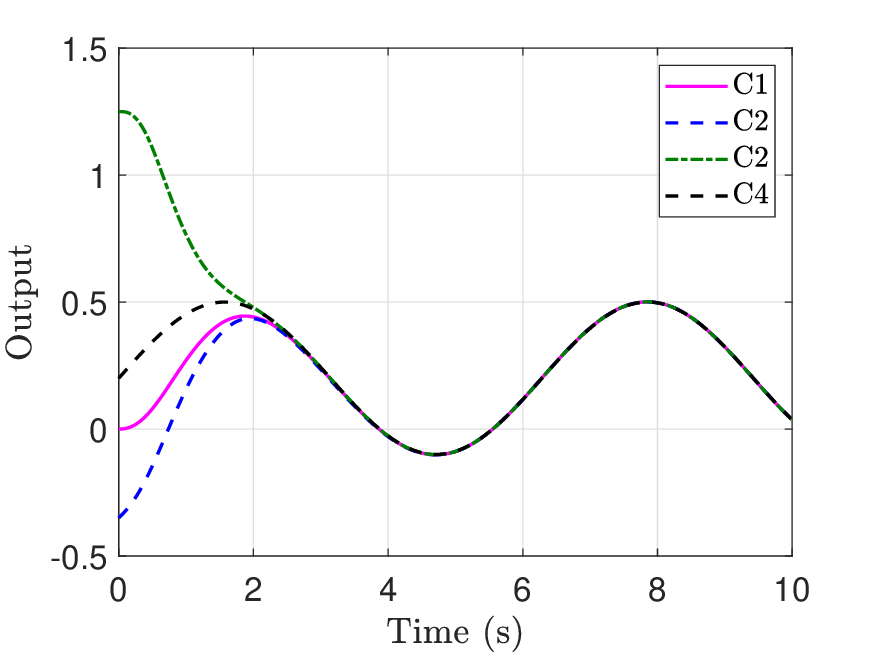}
     \caption{Output trajectories.}
    \label{fig:Bounded_Input_output}
\end{subfigure}%
\begin{subfigure}{0.25\linewidth}
    \centering
    \includegraphics[width=\linewidth]{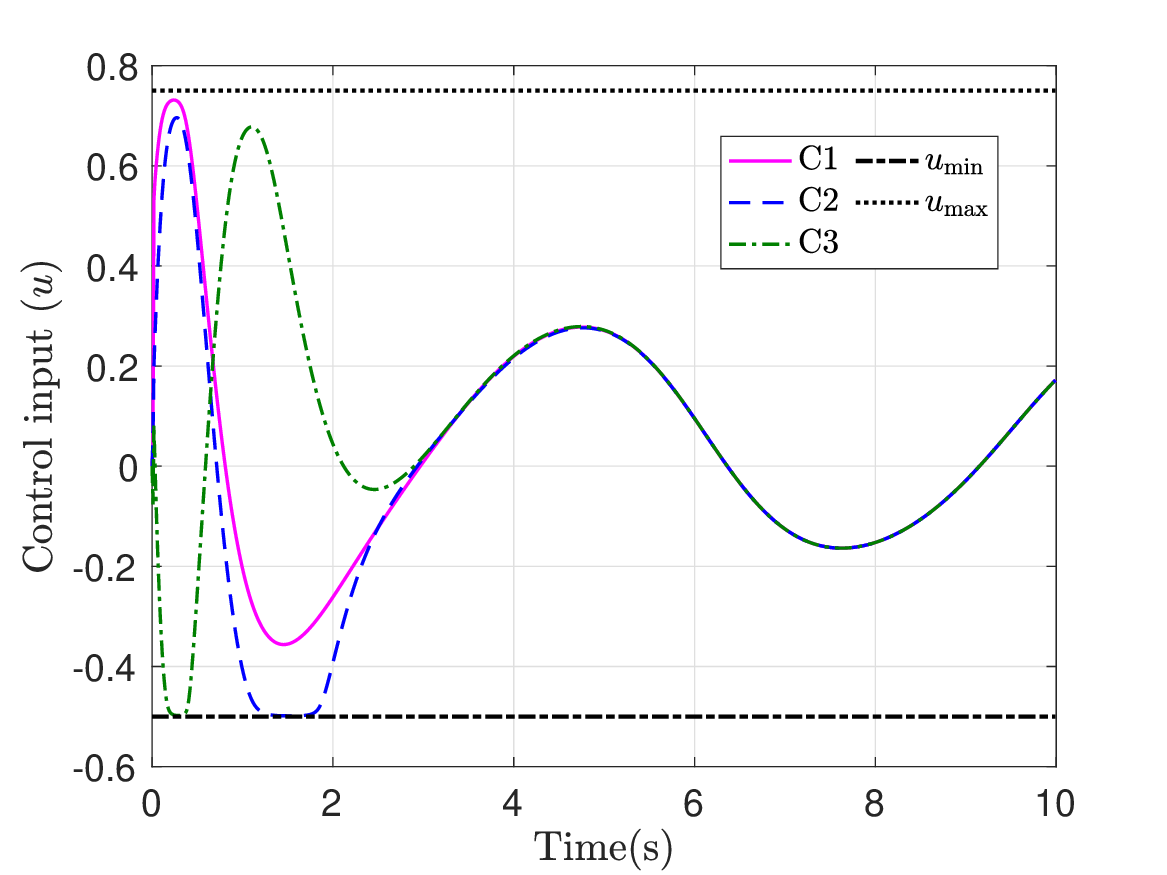}
    \caption{Control inputs.}
    \label{fig:Bounded_Input_input}
\end{subfigure}%
\begin{subfigure}{0.25\linewidth}
	\centering
    \includegraphics[width=\linewidth]{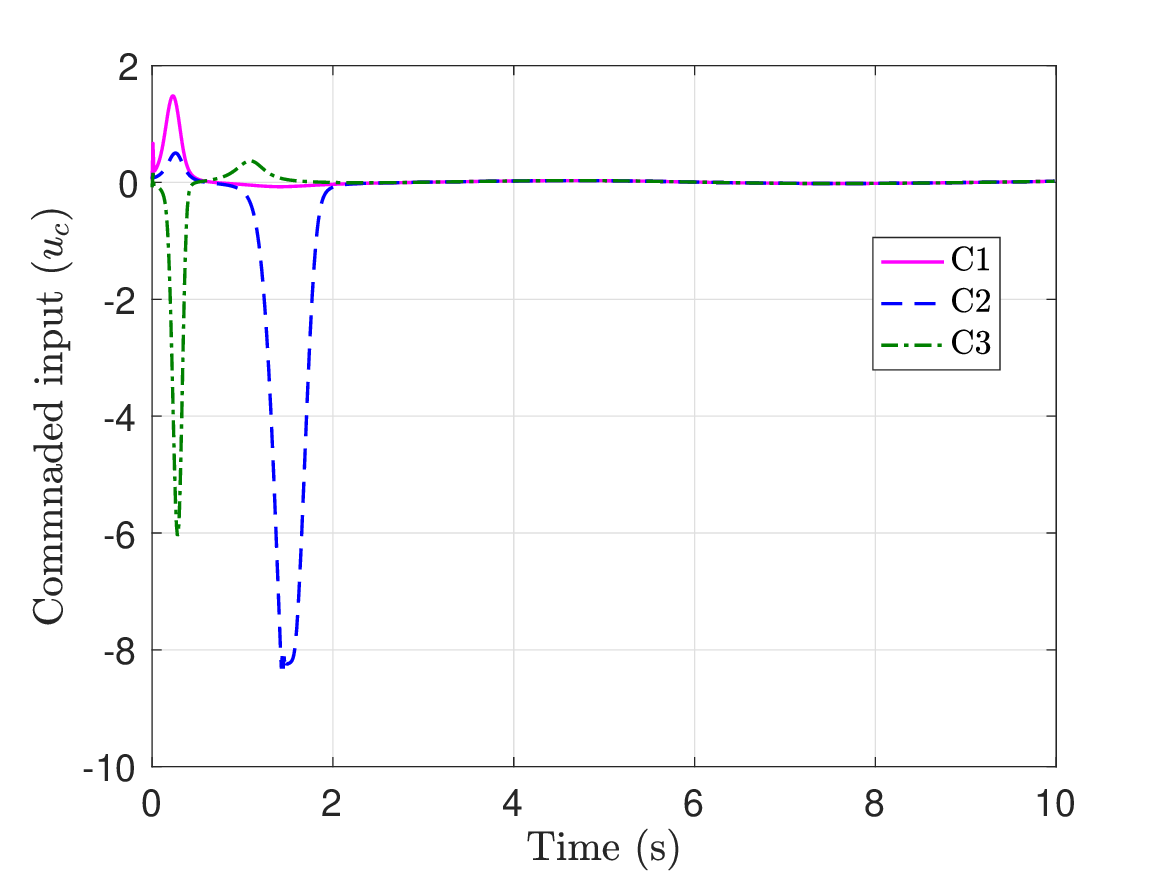}
     \caption{Commanded inputs.}
    \label{fig:Bound_Input_commanded_input}
\end{subfigure}%
\begin{subfigure}{0.25\linewidth}
    \centering
    \includegraphics[width=\linewidth]{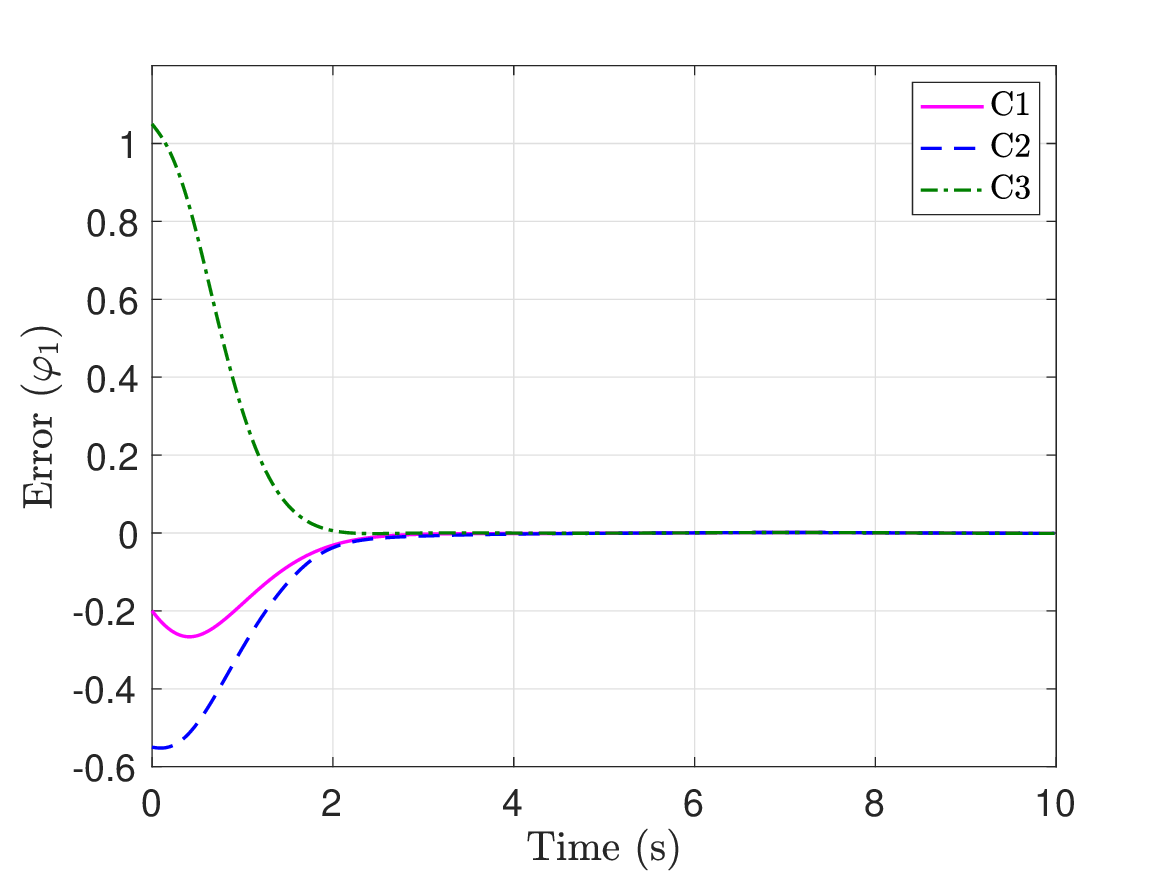}
    \caption{Tracking errors.}
    \label{fig:Bounded_Input_error}
\end{subfigure}
\caption{Output tracking with bounded input.}
\label{fig:Bounded_Input}
\end{figure*}
\begin{figure*}[!ht]
\centering
\begin{subfigure}{0.25\linewidth}
	\centering
    \includegraphics[width=\linewidth]{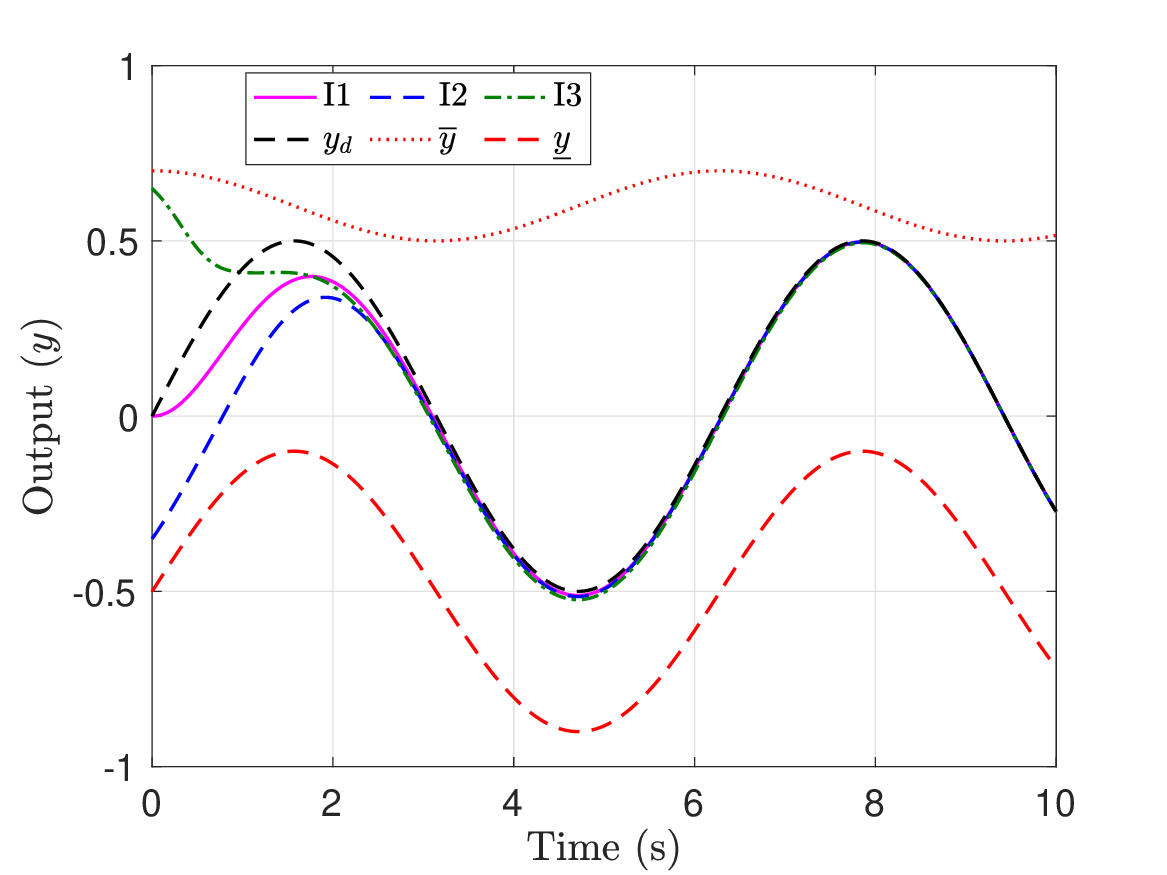}
     \caption{Output trajectories.}
    \label{fig:BI_Time_var_Output_output}
\end{subfigure}%
\begin{subfigure}{0.25\linewidth}
    \centering
    \includegraphics[width=\linewidth]{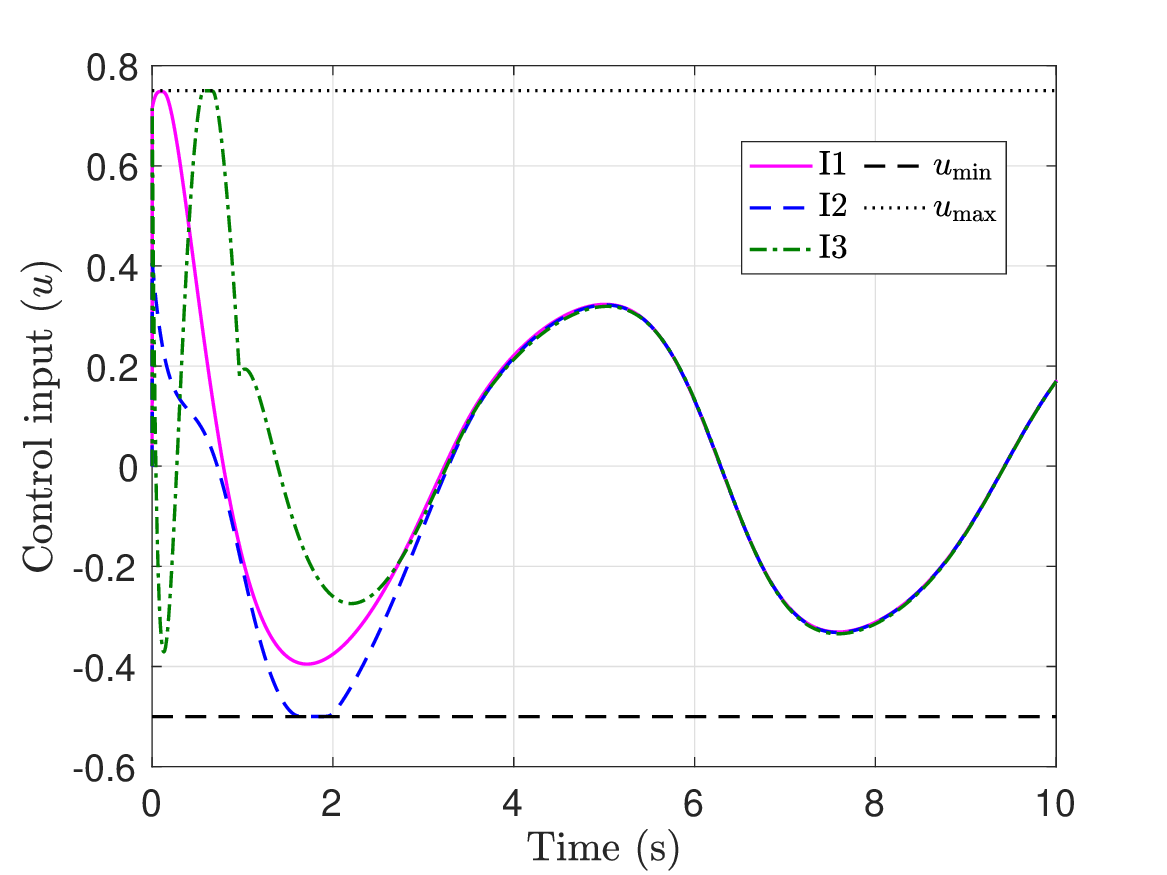}
    \caption{Control inputs.}
    \label{fig:BI_Time_var_Output_input}
\end{subfigure}%
\begin{subfigure}{0.25\linewidth}
    \centering
    \includegraphics[width=\linewidth]{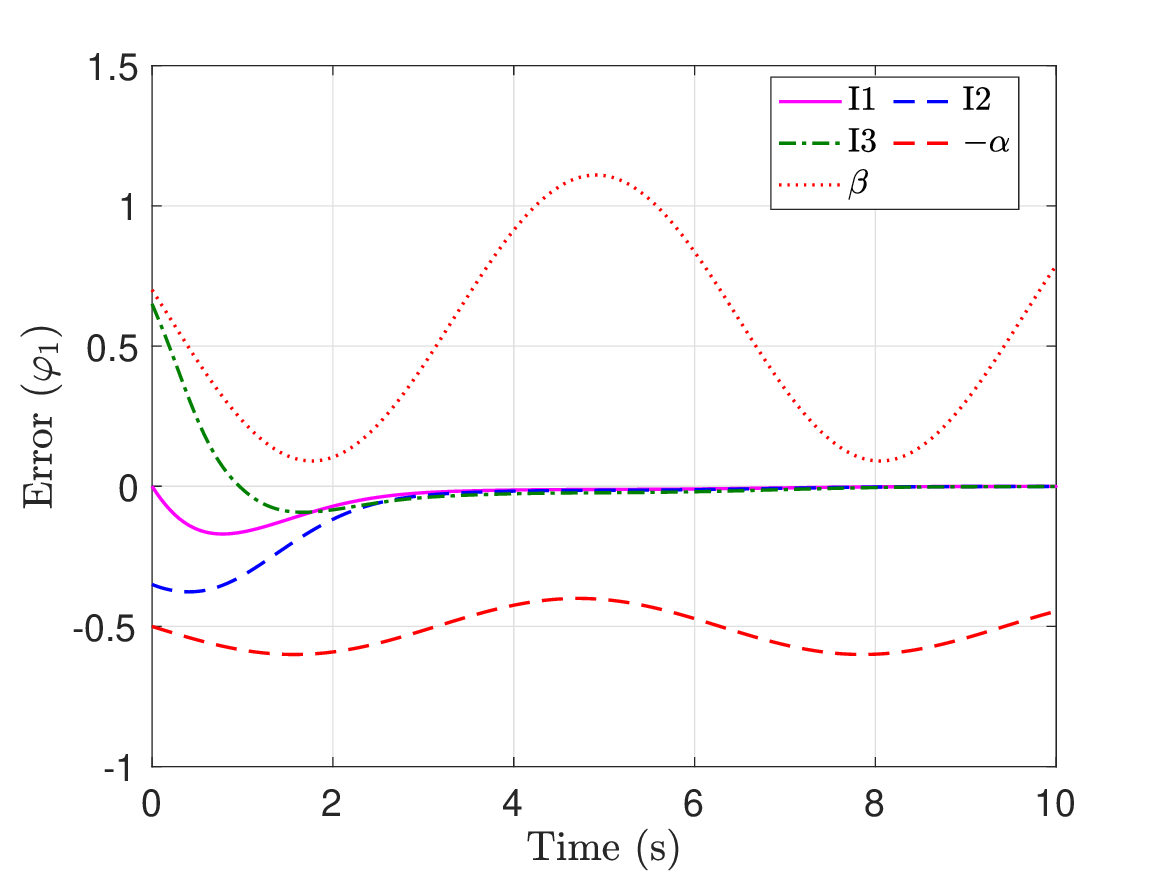}
    \caption{Tracking errors.}
    \label{fig:BI_Time_var_Output_error}
\end{subfigure}%
\begin{subfigure}{0.25\linewidth}
	\centering
    \includegraphics[width=\linewidth]{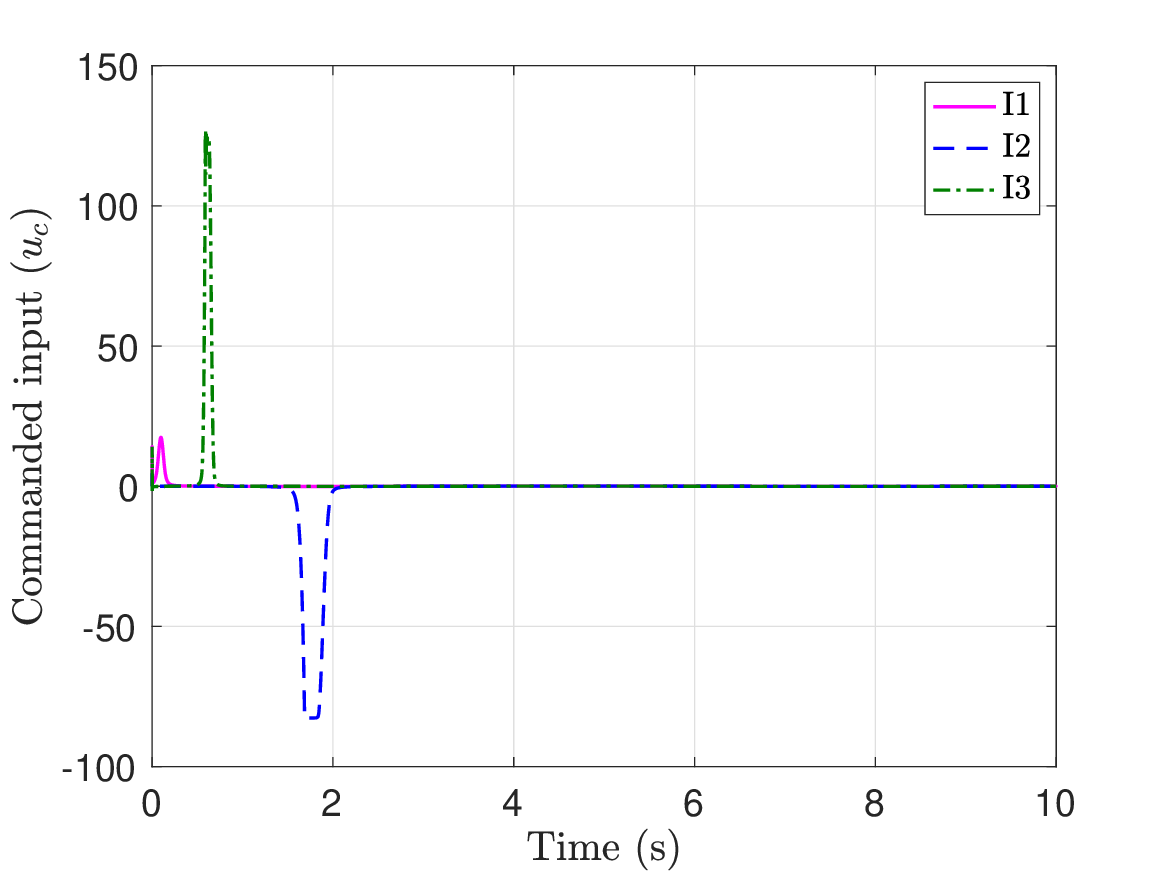}
     \caption{Commanded inputs.}
    \label{fig:BI_Time_var_Output_commanded_input}
\end{subfigure}
\caption{Constrained (time-varying) output tracking with bounded input.}
\label{fig:BI_Time_var_Output}
\end{figure*}
\begin{figure*}[!ht]
\centering
\begin{subfigure}{0.25\linewidth}
	\centering
    \includegraphics[width=\linewidth]{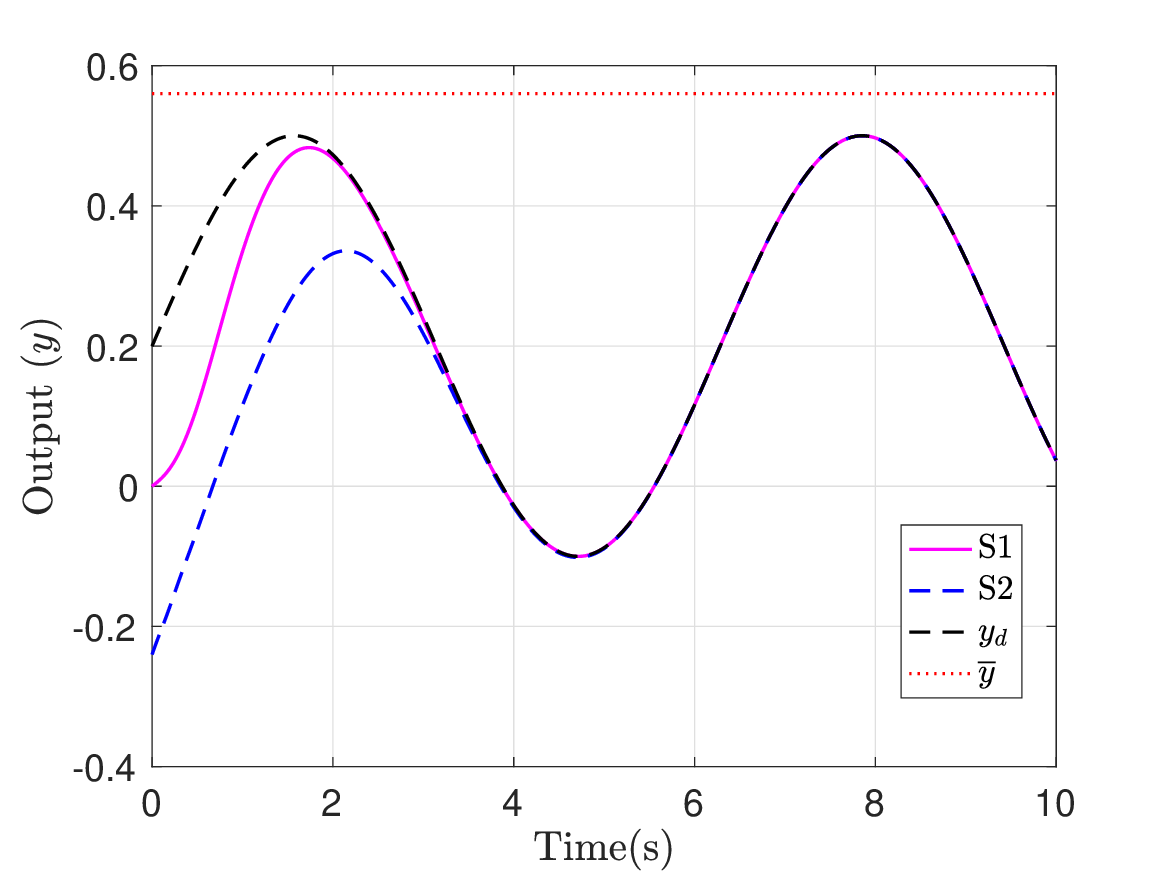}
     \caption{Output trajectories.}
    \label{fig:BI_Constant_Output_output}
\end{subfigure}%
\begin{subfigure}{0.25\linewidth}
    \centering
    \includegraphics[width=\linewidth]{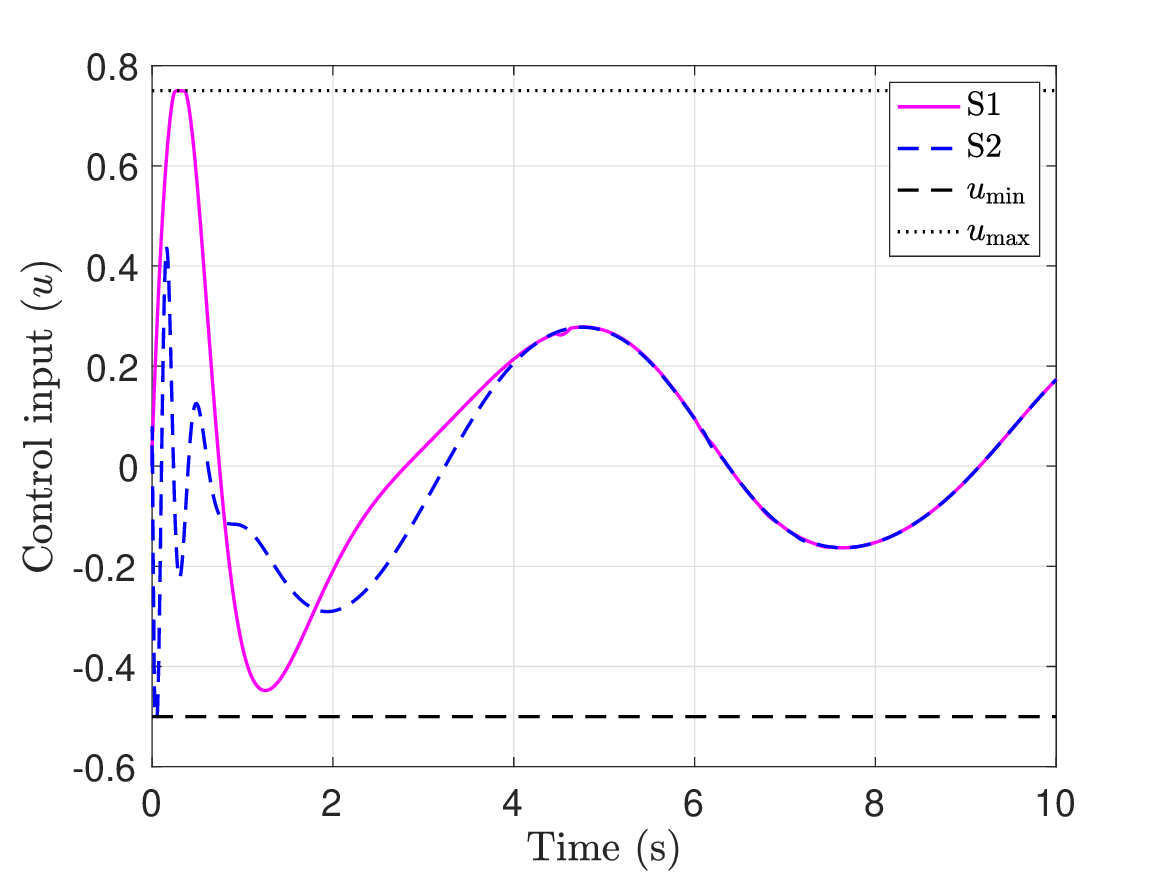}
    \caption{Control inputs.}
    \label{fig:BI_Constant_Output_input}
\end{subfigure}%
\begin{subfigure}{0.25\linewidth}
    \centering
    \includegraphics[width=\linewidth]{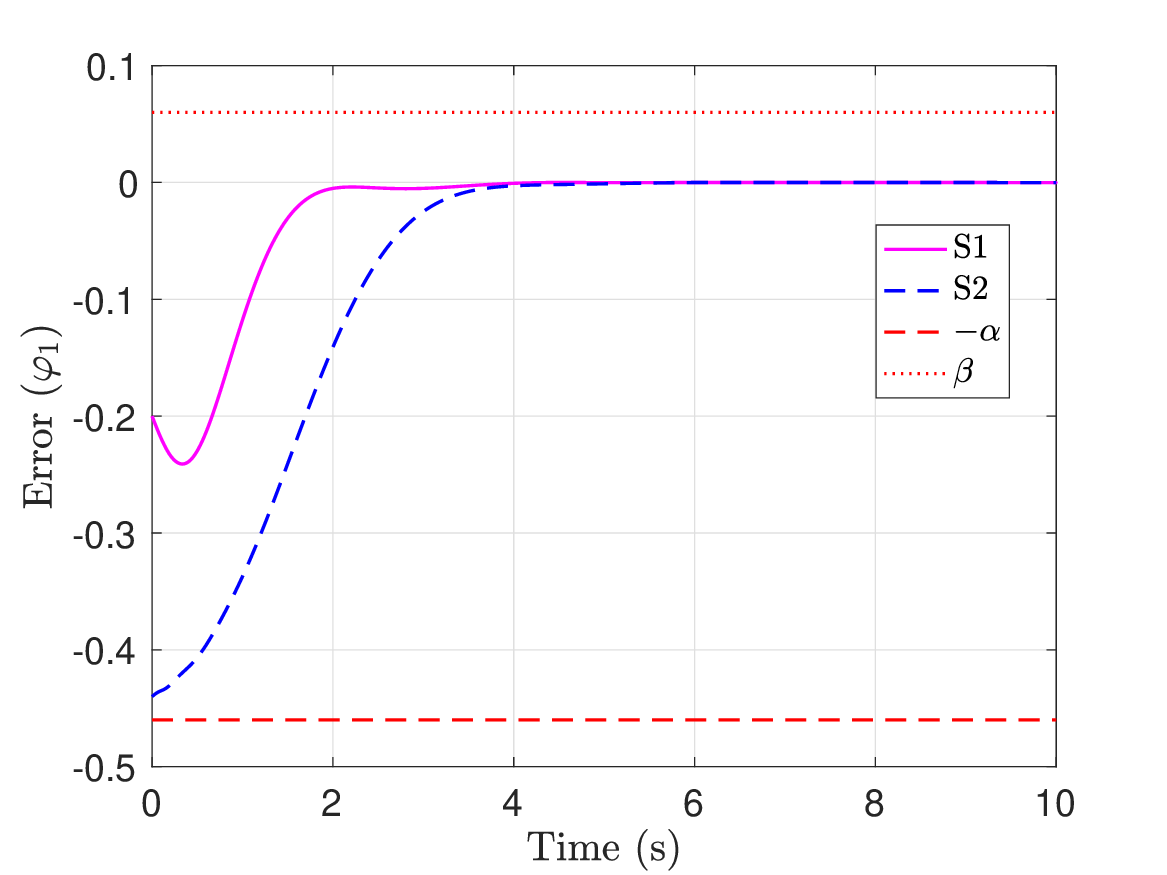}
    \caption{Tracking errors.}
    \label{fig:BI_Constant_Output_error}
\end{subfigure}%
\begin{subfigure}{0.25\linewidth}
	\centering
    \includegraphics[width=\linewidth]{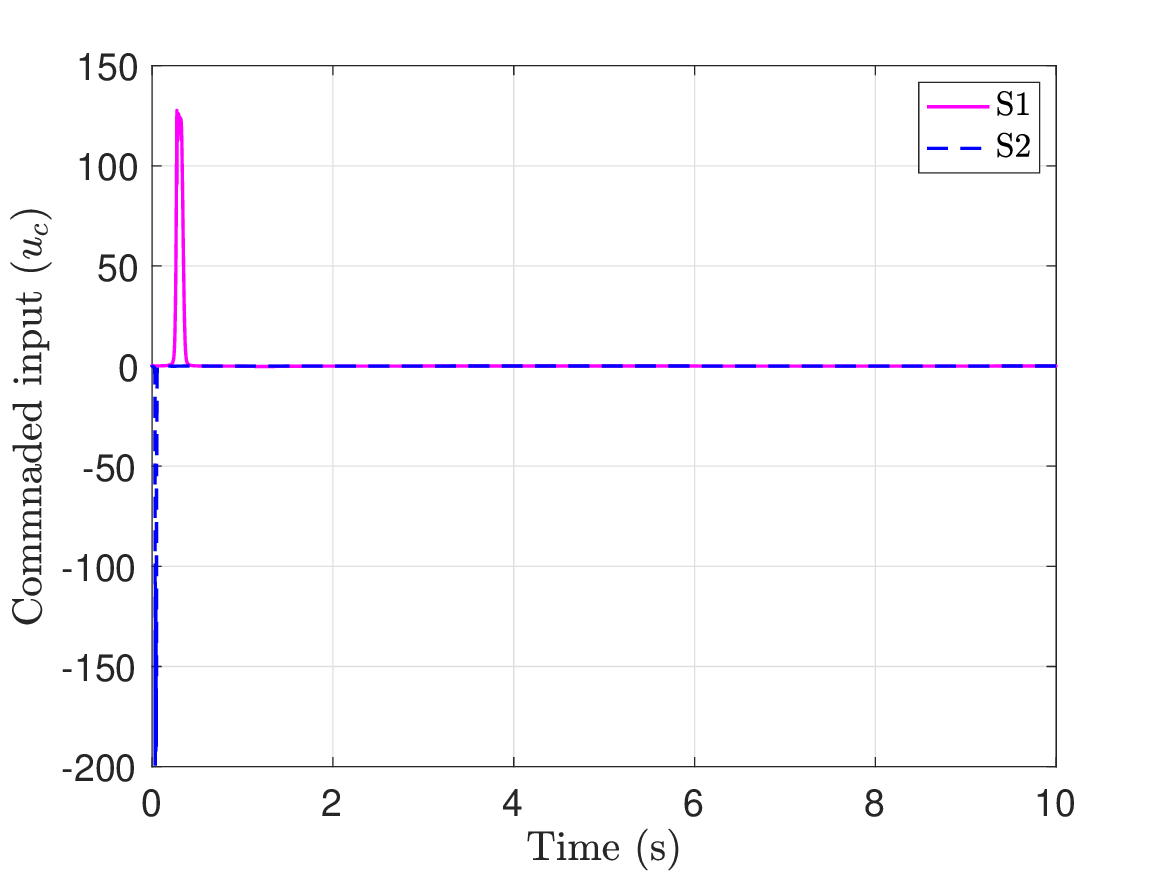}
     \caption{Commanded inputs.}
    \label{fig:BI_Constant_Output_commanded_input}
\end{subfigure}
\caption{Constrained (constant) output tracking with bounded input.}
\label{fig:BI_Constant_Output}
\end{figure*}
In what follows, we demonstrate the effectiveness of the proposed nonlinear control strategy through a series of scenarios involving output tracking under bounded input constraints and constrained output tracking with bounded inputs. We consider the following second-order nonlinear system as a prototype 
\begin{align*}
\dot{x}_{1} =&~ 0.1 x_{1}^2 + x_{2},\\
\dot{x}_{2} = &~0.1x_{1}x_{2}-0.2x_{1} + (1+x_{1}^2) u.
\end{align*}
The system is subject to asymmetric input constraints as defined in \eqref{eq:sat_def}, with lower and upper bounds chosen as $u_{\min} = -0.5$ and $u_{\max} = 0.75$, respectively. The other design parameters for the input saturation model are chosen to be $p_{1}=100$, $p_{2}=0.1$, and $\gamma=2$. The desired output trajectory is selected as $y_d=0.2 + 0.3 \sin(t)$. In the following control input plots, the dashed and dotted black lines denote the upper and lower bounds of the input, respectively. Likewise, in all output trajectory plots, the desired trajectory is represented by a black dashed line.

We first consider the scenario when the plant output has to track a time-varying desired trajectory subject to the given control constraints. The controller parameters are chosen as $k_1=k_2=k_3=2$. We consider three different initial conditions, which are represented by C1, C2, and C3 in the following plots. The initial conditions are represented by ordered pair ($x_1(0),x_{2}(0)$) and chosen to be $(0,0)$, $(-0.35,0.25)$, and $(1.25,-0.15)$, for C1, C2, and C3, respectively. With the given initial conditions and the design parameters, we show the performance of the proposed control strategy through Fig.~\ref{fig:Bounded_Input}. One may notice from Fig.~\ref{fig:Bounded_Input_output} that regardless of initial conditions, the output follows its desired time-varying trajectory. It can be observed from Fig.~\ref{fig:Bounded_Input_input} that the plant input remains within the prespecified bounds for all cases. Although the commanded control inputs (Fig.\ref{fig:Bound_Input_commanded_input}) may at times exceed these limits, they remain finite, thereby satisfying the sufficient condition of Theorem~\ref{thm:sat_model} that ensure the plant input stays remain confined to the predefined set. One may also notice from Figs.~\ref{fig:Bounded_Input_input} and \ref{fig:Bound_Input_commanded_input} that whenever the commanded control exceeds saturation limits, the plant input goes near the boundaries but never crosses them.

Next, we consider a scenario when the system's output is subjected to asymmetric time-varying output constraints, and the system's input is also subjected to asymmetric input constraints. The desired trajectory has been chosen to be the same as the previous case, that is, $y_{d}= 0.2 + 0.3 \sin{t}$. The asymmetric time-varying output constraints are chosen to be $\overline{y}(t)=0.6+0.1\cos(t)$ and $\underline{y}(t)=-0.5+0.4\sin(t)$. These upper and lower bounds are denoted by dotted and dashed red lines, respectively, in the following trajectory plot. Here, we choose different initial conditions as ($0,0$), ($-0.35,0.35$), and ($0.65,-0.32$), which are represented by I1, I2, and I3, respectively, in the following plots. The time-varying bounds on the output tracking error $\varphi_{1}(t)$, that is, $\beta(t)$ and $\alpha(t)$ are denoted by dotted and dashed red color lines in the following error plots. All other simulation set-up is kept the same as before. Under the proposed control strategy, we show the performance of the second-order system using Fig.~\ref{fig:BI_Time_var_Output}. It can be observed from Fig.~\ref{fig:BI_Time_var_Output_output} that the system output tracks its desired trajectory while respecting the time-varying constraints. The commanded and actual control inputs, depicted in Figs.~\ref{fig:BI_Time_var_Output_commanded_input} and \ref{fig:BI_Time_var_Output_input}, respectively, demonstrate behavior similar to the previous case, affirming the effectiveness of the proposed control strategy. It can be observed from Fig.~\ref{fig:BI_Time_var_Output_error} that the tracking errors also remain within the specified time-varying bound throughout the engagement.

We now consider a scenario when the plant is subjected to constant output constraints with asymmetric input constraints. For this case, the initial condition is chosen to be ($0,0.1$) and ($-0.24,0.36$), denoted by S1 and S2 in the following plots. While keeping the other engagement setting the same as for the time-varying output constraints case, we now show the performance of the proposed control strategy through Fig.~\ref{fig:BI_Constant_Output}. For ease of visualization, only the upper bounds are shown in the trajectory plot since the output never reaches its lower bound during the given engagement. One can notice that the plant output converges to the desired trajectory while adhering to the constant output and asymmetric input constraints by exhibiting behavior similar to the time-varying constraint case.

\section{Conclusions}\label{sec:conclusions}
In this work, we addressed the problem of output tracking for a class of nonlinear systems under input and output constraints. Two nonlinear control strategies were developed to ensure that the system output tracks the desired trajectory while satisfying strict actuator magnitude limits and remaining within a prescribed safe output set. The proposed formulation guarantees the boundedness of all closed-loop signals and provides a mathematically rigorous framework for simultaneously handling both actuator and output constraints. This work is expected to provide a significant step toward constraint-aware nonlinear control design, with implications for safety-critical systems in aerospace, robotics, process control industries, and various other control applications where constraints are prevalent. Incorporating model uncertainties and external disturbances could be some of the interesting future research directions of the current work.
\bibliographystyle{ieeetr}
\bibliography{ref_input_sat.bib} 
\end{document}